\documentclass{jaa}
\usepackage{natbib}
\usepackage{graphicx}
\usepackage{amssymb}
\usepackage{amsmath}
\usepackage{array}
\usepackage{lineno}
\usepackage{multirow}
\usepackage{subcaption}
\usepackage{amsfonts}
\usepackage{bm}
\usepackage{bigstrut}
\usepackage{aas_macros}
\usepackage{slashbox}

\begin{document}\sloppy

\title{Inference of cosmological models with principal component analysis}


\author{Ranbir Sharma\textsuperscript{1,*} and H K Jassal\textsuperscript{1}}
\affilOne{\textsuperscript{1}Indian Institute of Science Education and Research Mohali, Punjab, 140306, India.\\}


\twocolumn[{

\maketitle

\corres{ranbirsharma0313@gmail.com}


\begin{abstract}
Determination of cosmological parameters is a major goal in cosmology at present.
The availability of improved data sets necessitates the development of novel statistical tools to interpret the inference from a cosmological model.  
In this paper, we combine the Principal Component Analysis (PCA) and Markov Chain Monte Carlo (MCMC) method to infer the parameters of cosmological models. We use the No U-Turn Sampler (NUTS) to run the MCMC chains in the model parameter space.
After determining the observable by PCA, we replace the observational and error parts of the likelihood analysis with the PCA reconstructed observable and find the most preferred model parameter set.
As a demonstration of our methodology, we assume a polynomial expansion as the parametrization of the dark energy equation of state and plug it in the reconstruction algorithm as our model.
After testing our methodology with simulated data, we apply the same to the observed data sets, the Hubble parameter data, Supernova Type Ia data, and the Baryon Acoustic oscillation data.
This method effectively constrains cosmological parameters from data, including sparse data sets.
\end{abstract}


\keywords{Cosmology---(dark energy equation of state reconstruction)---(Principal Component Analysis)---(correlation coefficient)}
}]


\doinum{12.3456/s78910-011-012-3}
\artcitid{\#\#\#\#}
\volnum{000}
\year{0000}
\pgrange{1--}
\setcounter{page}{1}
\lp{1}

\section{Introduction}
\label{introduction}

Observational evidence of the acceleration of the Universe marked the beginning of a new era in Cosmology.
It is well established that the current expansion of the Universe is accelerating, and
an explanation for the current acceleration is done by way of introducing the Dark Energy(DE) term in the Einstein equation.  
Dark energy is described by its equation of state parameter (EoS)
$w=-P^{\prime} / \rho^{\prime}$,  where $\rho^{\prime}$ is the energy density and $P^{\prime}$ is its pressure contribution.  
It is still unknown whether dark energy is a cosmological constant
\citep{Carroll1992,Carroll:2000fy,Turner:1998ex,Padmanabhan:2002ji} or
a time-evolving entity \citep{Peebles:2002gy,Copeland:2006wr}. 
The $\Lambda CDM$ ({\it cosmological constant} and {\it cold dark
  matter}) model corresponds to the dark energy equation of state value $w=-1$, whereas in the case of time-evolving dark energy,  the dark energy equation of state parameter varies with time and
can assume different values of $w$
\citep{Padmanabhan:2002ji,Peebles:2002gy,Carroll1992,Weinberg1989,Coble1997,Caldwell1998,
Sahni2000,Ellis2003,Linder2008,Frieman2008,Albrecht2006,licia2008,nesseris2020}.
Various models based on scalar, canonical, and non-canonical fields have
been proposed to overcome different problems of $\Lambda CDM$ model
\citep{Ratra:1987rm,Linder:2006sv,Caldwell:2005tm,Linder:2007wa,Huterer:2006mv,Zlatev:1998tr,
Copeland:1997et,PhysRevD.66.021301,Singh:2018izf,Bagla:2003prd,Tsujikawa:2013,Rajvanshi:2019wmw,
Chevallier:2000qy}.
The discrepancies in $H_0$ measurements and its implication in cosmological 
model selection is discussed in \cite{Banerjee2021prd} and  \cite{Lee2022jcap}.
The last two decades have also marked the era of precision Cosmology.
Cosmological parameters are measured to high precision utilizing the availability of new data-sets \citep{Aghanim:2018eyx,Chevallier:2000qy,Archana:2018}.

Maximum Likelihood Estimation analysis(MLE) is the most commonly used technique in
cosmological parameter estimation\citep{Singh:2018izf,Jassal:2009ya,Nesseris:2004prd,Nesseris:2005prd,Nesseris:2007,Archana:2018}. 
The increasing availability of the observational data-set has tightened the constraints on the parameters of  theoretical
models \citep{Chevallier:2000qy,Linder:2002et,Jassal:2004ej,Gong:2006gs,Mukherjee:2016eqj,2018PhRvD..98h3501V,
DiValentino:2017zyq,licia2020_1,licia2020_2,licia:2013}.
Though it is crucial to determine the theory parameters, we have the
observational data dependencies in the core of these methods, and new
data sets reject or accept a particular model with quantified precision.
Methods like the Principal Component Analysis (PCA) enable us to determine the
functional form of the observable of a data-set in a model 
independent, non-parametric manner \citep{Huterer2003prl,Huterer2005,Zheng:2017ulu,Ishida2011is,Crittenden:2005wj,clarkson2010prl,Miranda:2018,
rah_2020,Hart:2019,Nesseris:2013PhRvD,Nair2013,Hojjati:2012prd,Hart2022}. 
PCA is a multivariate analysis that gives the form of cosmological quantities as a function of redshift \citep{Huterer2003prl,clarkson2010prl, Huterer2005, Zheng:2017ulu, rah_2020}.
In a previous work \citep{rah_2020}, we combined PCA and Correlation Coefficient Calculation to give the analytical, functional form of the observable quantity when observational data-sets
are given as input. 
The method is efficient in fitting the observable; the caveat however is that the derived cosmological parameters like the dark energy equation of state parameter are not determined very efficiently.
The problem arises due to the non-linear dependency of the dark energy parameter to the observational quantity at hand, for instance, the Hubble parameter and the distance modulus.
To circumvent this problem, we incorporate the Markov Chain Monte Carlo method with PCA reconstruction to derive the equation of state parameters for dark energy and other cosmological parameters.
The equation of state parameter is derived by searching for the model that best describes the functional form  of the observable determined by the observational data.
For the Monte Carlo method, we use the No U-Turn Sampler which is  a variant of the Hamilton Monte Carlo method. 
In this analysis, we show that the constraints on the dark energy equation of state parameters are consistent with the constraints obtained from other methods. 

This paper is structured as follows. 
In section \ref{sec_2}, we give a brief review of background cosmology, we describe the reconstruction algorithm along with the No U-turn sampling, followed by section \ref{sec_3}  where we describe the results of our algorithm.
We describe the distinguishing features of our methodology in section \ref{sec_new}. 
In section \ref{sec_4} we conclude by summarising the main results of this paper.


\section{Reconstruction Methodology} \label{sec_2}

In this section, we first discuss the methodology of the Principal Component Analysis reconstruction \citep{rah_2020}  and the modification to the algorithm. 

\subsection{Reconstruction of the functional form of Hubble parameter, distance modulus and angular scale in terms of redshift}

For a spatially flat Universe, composed of dark energy and non-relativistic matter,  the  Hubble parameter  is given by,

\begin{equation}
H(z) = H_0\left[ \Omega_m (1+z)^3 + \Omega_{DE} e^{3 \int^z_0 \frac{1+w(z')}{1+z'}dz'} \right]^{1/2} \label{eq:FRW}
\end{equation}

The dark energy equation of state parameter $w(z) = P^{\prime}/\rho^{\prime}$ can be written  as

\begin{equation}
w(z) = \sum_{i=1}^{m} \alpha_{(i - 1)} \mathcal{F}(z)^{(i - 1)}, \hspace{0.5cm}  \mathcal{F}(z) = \frac{z}{(1 + z)} \label{eq:param_eos} 
\end{equation}

where $H_0$ denotes the present-day value of the  Hubble parameter and $\Omega_m$, $\Omega_{DE}$ are 
the density parameters for matter and dark energy, respectively.  
In eqn(\ref{eq:param_eos}),  $m=2$ corresponds to the Chevallier-Polarski-Linder(CPL) parameterization \citep{cpl0, cpl1} given by,
$w(z) = w_0 + w' z /(1+z)$, $w_0$ and $w'$ being the present-day values of the equation state parameter and its derivative, respectively.
The equation gives the Taylor series expression of the dark energy equation of state parameter in terms of $(1 - a)$, where $a$ is the scale factor. 
 
From the functional form of the Hubble parameter, we can reconstruct the dark energy equation of state parameter $w(z)$.
Differentiating eqn (\ref{eq:FRW}) we get, 
\begin{equation}
    w(z) = \frac{3 h^2  - 2 (1 + z) h h^{\prime} }{3 h_0^2 (1 + z)^3 \Omega_m - 3 h^2},
    \label{wToh}
\end{equation}
where $h$ is the reduced Hubble parameter given by H(z)/$100 ~km ~s^{-1} Mpc^{-1}$.

The luminosity distance $d_L(z)$ is given by,
\begin{equation} \label{eq:d_L}
d_L(z) = \frac{c}{H_0}(1+z)\int_0^z d_H(z')dz'
\end{equation}

where $d_H$, from eq(\ref{eq:FRW}) is, 
\begin{equation} \label{eq:d_H}
d_H(z) = \left(\Omega_m (1+z)^3 + \Omega_x e^{{3\int_0^z \frac{(1+w(z'))dz'}{(1+z')}}}\right)^{-1/2}
\end{equation}
and is related to the distance modulus as  
\begin{equation} \label{eq:mu}
\mu(z) = 5 \log{\left(\frac{d_L}{1Mpc}\right)} + 25
\end{equation}
We use the same expression of eqn(\ref{eq:param_eos}) for the EoS parameter to express $\mu(z)$. 


Since $D(z) = (H_0 / c) (1+z)^{-1} d_L(z) $, the equation of state
parameter in terms of distance is given by
\begin{equation} \label{eq:EoS_SNIa}  
w(z) = \frac{2(1+z)D'' +3D'}{3D'^3 \Omega_m (1+z)^3 -3D'}
\end{equation}

The Baryon Acoustic Oscillation (BAO) angular scale $\theta_{b}$ is defined in terms of the angular diameter $D_A$ as,

\begin{equation} \label{eqn:bao}
    \theta_b = \frac{r_{drag}}{(1 + z) D_A}.
\end{equation}

Here, $r_{drag}$ is the sound horizon at the drag epoch.

Following the reconstruction method of \cite{rah_2020}, we start by calculating the functional form of the reduced Hubble parameter
$h(z)$ and distance modulus $\mu(z)$ directly from the data-set, using Principal Component Analysis.  
The observable of the given data-set is expressed as a polynomial over an initial basis function,
    which creates a coefficient space. 
    The dimension of the coefficient space is the same as the number of terms in the 
    initial basis function.  
 We select different patches in the coefficient space and do a 
    $\chi^2$ calculation on each patch.
   For each patch, we get a minimum value of $\chi^2$.
   From these minimum $\chi^2$ values of each patch, we create the PCA data-matrix($\mathcal{D}$).
We then calculate covariance matrix $\mathcal{C}$ of $\mathcal{D}$, from 
    which the eigenvector matrix $\mathcal{E}$ is calculated.
    $\mathcal{E}$ is used to diagonalize $\mathcal{C}$ and omit the linear correlation of the data matrix.
    It also creates a new set of basis functions.
 The observables are finally expressed in terms of the final basis function. 
 With the help of these new basis functions, we create the new data-matrix $\mathcal{D}'$. 
    To select the value of the final basis number $M$, we compare the correlation matrix of $\mathcal{D}$ and $\mathcal{D'}$.
    Comparison of the correlation matrix also helps us to choose the better initial basis variable.

If the initial basis function is given by 
\begin{displaymath}
G=(f_1(z), f_2(z), ...., f_N(z)), 
\end{displaymath}
with $f_i(z) = f(z)^{(i-1)}$, the initial expression of the observable $\xi$ in terms of the 
independent variable $z$ is given by,
\begin{equation}
\xi_{ini}(z) = \sum_{i=1}^{N} b_i f(z)^{(i -1)}
\end{equation}
The value of $N$ is the number of terms in the polynomial expression of $\xi_{ini}(z)$; it is also the dimension of coefficient space $\vec{b}$.
The correlation coefficient calculation determines the value of $N$ \citep{M_G_Kendall_1938,rah_2020}. 
This value must be large enough that the function can capture most of the features from the observed data-set. 
To select the value of $N$, we calculate Pearson, Spearman, and Kendall correlation coefficients 
for the data-matrix $\mathcal{D}$ \citep{M_G_Kendall_1938,kreyszig2011advanced}. 
The Pearson correlation coefficient gives the linear correlation that exists in the data-set. 
On the other hand, Spearman and Kendall correlation coefficients give the non-linear correlations of 
the data-set.
For the Spearman correlation coefficient, we calculate the rank of the data-set. 
We arrange the ranks according to the numerical value; that is, we give rank 1 to the highest numerical value of the PCA data-set, rank 2 to the second highest, and so on.
The Spearman correlation coefficient is the Pearson correlation coefficient of the rank variable of the data-set. 
Spearman correlation gives information about whether the dependent and independent variables are monotonically increasing or decreasing.
For the Kendall correlation coefficient, we find the concordant and disconcordant pairs. 
It gives the ordinal association between the variables \citep{kreyszig2011advanced, M_G_Kendall_1938}.

We choose the smallest value of $N$ from the set of which the PCA data matrix gives us a higher value of Pearson Correlation coefficient
compared to the Spearman and Kendall correlation coefficients.  
Only if the expression of the observable $\xi(z)$ in terms of the polynomial is exact, there would 
no correlation between the coefficients of the polynomial expression.
Our motive is to break the correlation of the coefficient and obtain the polynomial expression of $\xi_{ini}(z)$ as closely as possible to the actual $\xi(z)$.
After the reduction of the higher order Principal Component(PC)s, the number of the terms in the polynomial of $\xi_{ini}(z)$ is $M$.
The final functional form of the observable is, 
$$\xi_{pca}(z) = \sum_{i=1}^{M} \kappa_i u_{i}(z) $$ 
where, $(u_1(z), u_2(z), ...., u_M(z))$ and $U = G \mathcal{E}$. After applying PCA, the dimension of the coefficient space $\vec{\kappa}$ is $M$.

In the earlier work, we have shown that a derived approach where the Principal Component Analysis obtains the observable and then reconstruction of dark energy equation of state parameter is an efficient method to reconstruct dark energy model than directly attempting to reconstruct it.
Also, while  we can reconstruct the Hubble parameter $h(z)$ very well with PCA, 
the presence of a differentiation term in the equation given by eqn (\ref{wToh}) 
which relates the EoS with $h(z)$ increases the errors in the reconstruction of $w(z)$. 

We address this problem by suggesting a modified approach to bypass the differentiation in calculating EoS from the PCA reconstructed  Hubble parameter, distance modulus function, and angular scale of BAO. 
This has been done by combining PCA with the Maximum Likelihood Estimation technique (MLE), using Markov Chain Monte Carlo (MCMC) to search for the best fit dark energy model to the PCA reconstructed Hubble parameter, distance modulus, and angular scale.
We replace the observational part of the MLE calculation with best-fit curve of $h(z)$, $\mu(z)$, and $\theta_b(z)$ as a function of redshift obtained via PCA.
This method omits the dependencies on the number of observational data points. 
This analysis gives us the machinery to produce the most probable value of the model
parameters by constraining the theory with reconstructed PCA data.
The errors are the error-functions created from the covariance
matrix of PCA data-matrix \citep{Huterer2003prl,rah_2020,clarkson2010prl}.
 The error comprises the eigenvalues and eigenfunctions of the covariance matrix \citep{Huterer2003prl, rah_2020, clarkson2010prl}. 
The eigenvalues of the covariance matrix quantify the error in the reconstruction of the 
observable $\xi(z)$. 
If $\lambda_i$ are the eigenvalues of the covariance matrix $\mathcal{C}$, then the error associated 
with each of the components is $\sigma(\alpha_i) = \lambda_i ^ {1/2}$. 
For $M$ number of final terms, we have the final error as, 

\begin{equation} \label{eqn:error_pca}
\sigma(\xi(z_a)) = \left[ \sum_{i=1}^{M} \sigma^2(\alpha_i) e^2_i(z_a) \right] ^ {1/2} 
\end{equation}
Eqn(\ref{eqn:error_pca}) gives the error function for a particular reconstructed curve, and we  have the error as a function of redshift \citep{Huterer2003prl, clarkson2010prl}.

\begin{figure*}
        \includegraphics[scale=0.18]{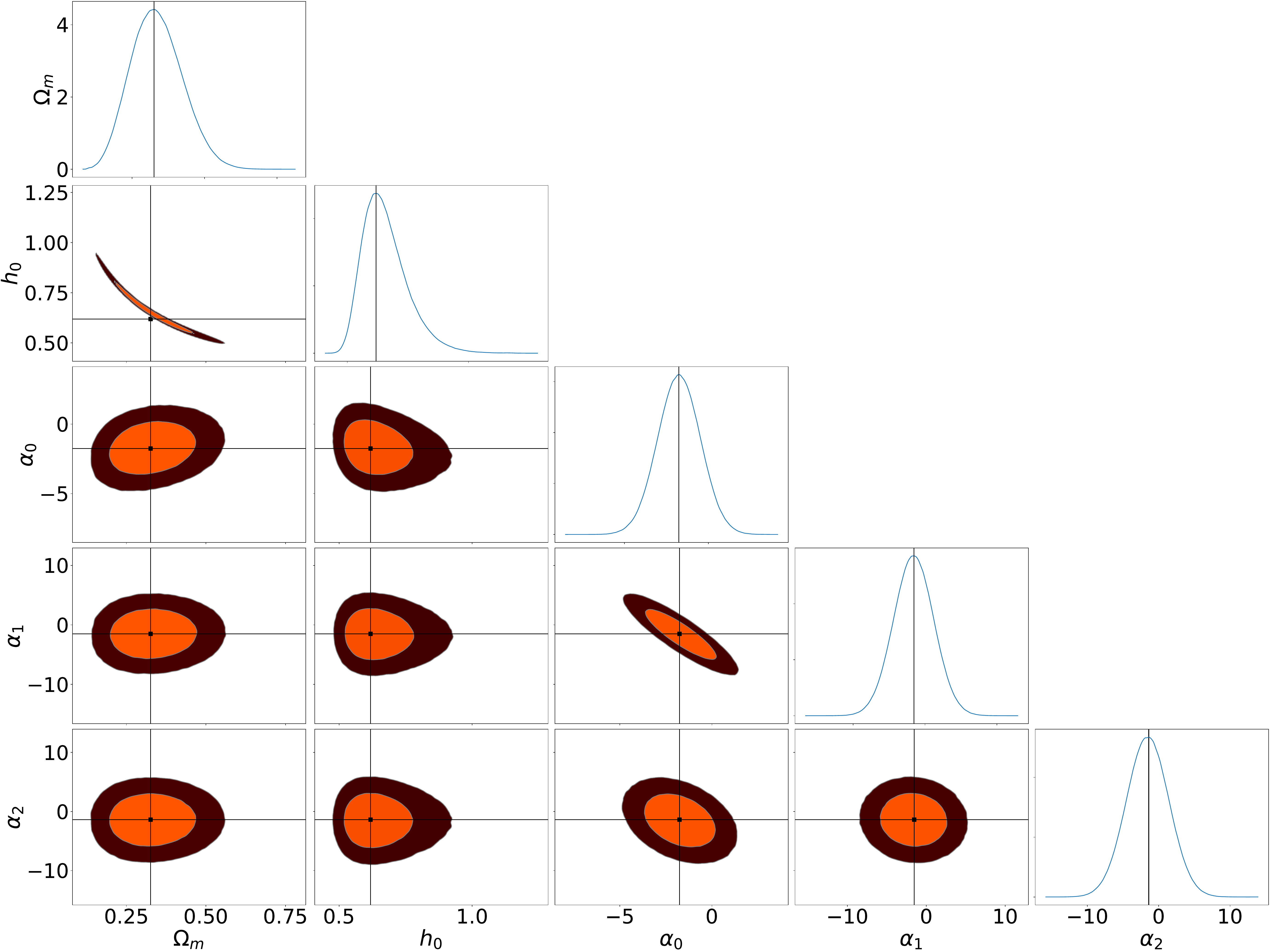}
    \caption{The plot shows the $1 \sigma$ and $2 \sigma$ contours for all the parameters along with their marginal probability density plots for the case of the simulated data-set.
        The first plot of every column is the marginal probability density plot, which give the maximum evidence for the real Hubble parameter data-set.}
\label{fig:simu_cont}
\end{figure*}

\begin{figure*}
\begin{center}
\includegraphics[scale=0.25]{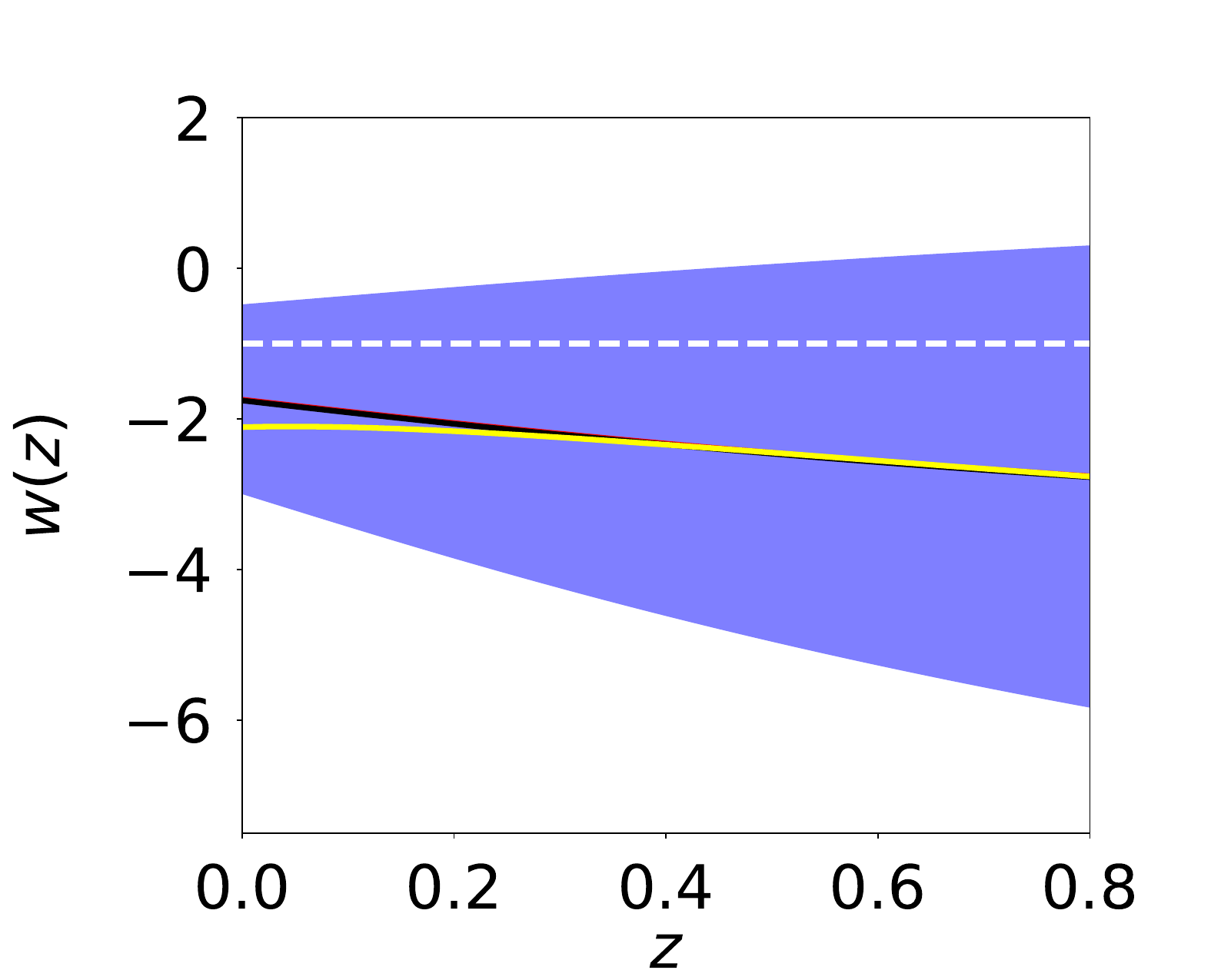}
\includegraphics[scale=0.25]{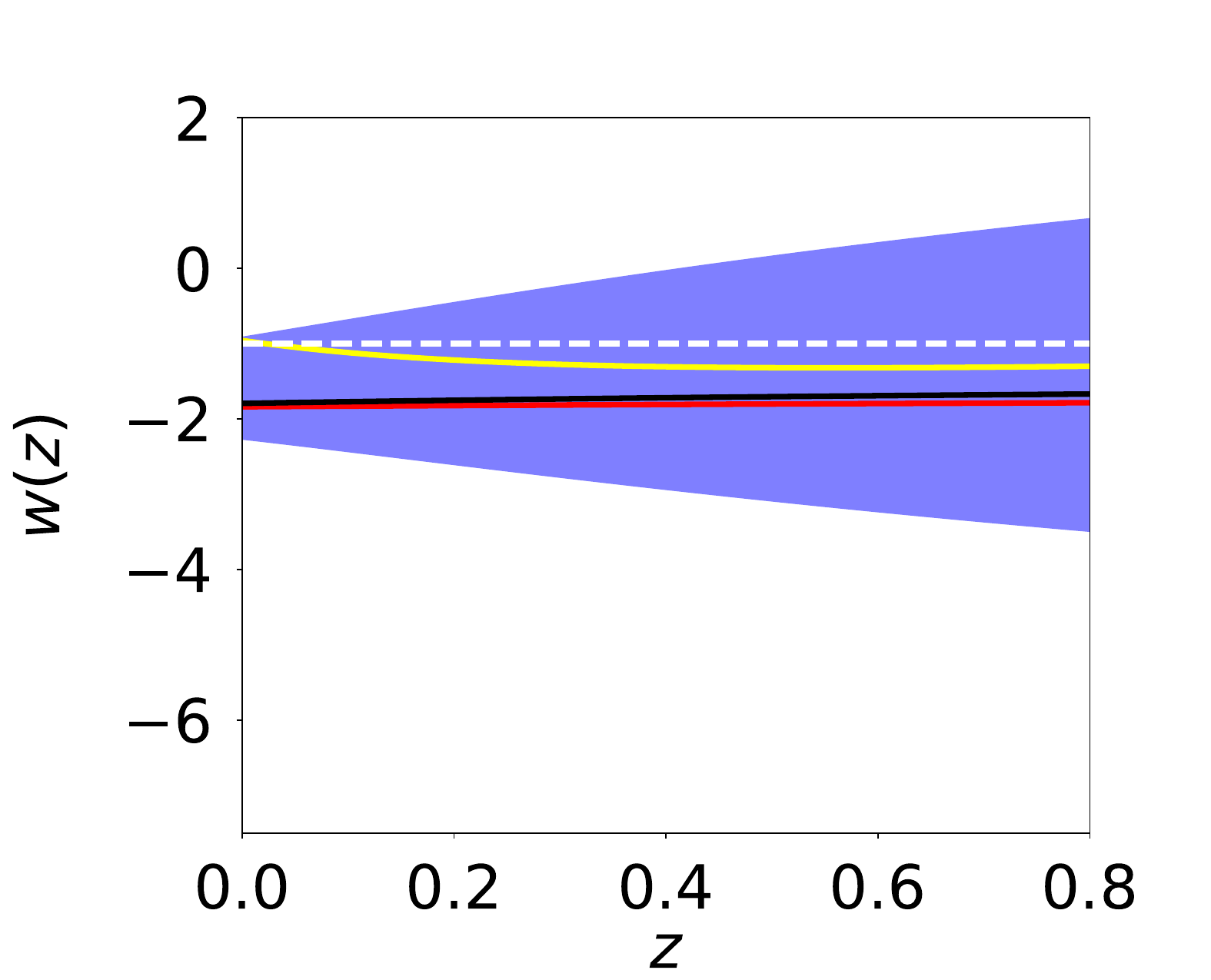}
\end{center}
\caption{In this figure, we show the  $68 \%$ confident level range for the equation of state parameter $w(z)$.
The figure on the left is for the simulated Hubble parameter data set, and the right is for the real data set.
The value of observational points $n_d$ and sample points $M_s$ are 600 and 800000, respectively.
The black, red, and yellow curves are the median, mean, and mode of the posterior density function.
The cosmological constant model is consistent with the data and is denoted by the dashed line. 
For both plots, the mean and median lines overlap.}
\label{fig:EoS_ranges_simu_and_real}
\end{figure*}

\begin{figure*}
        \includegraphics[scale=0.18]{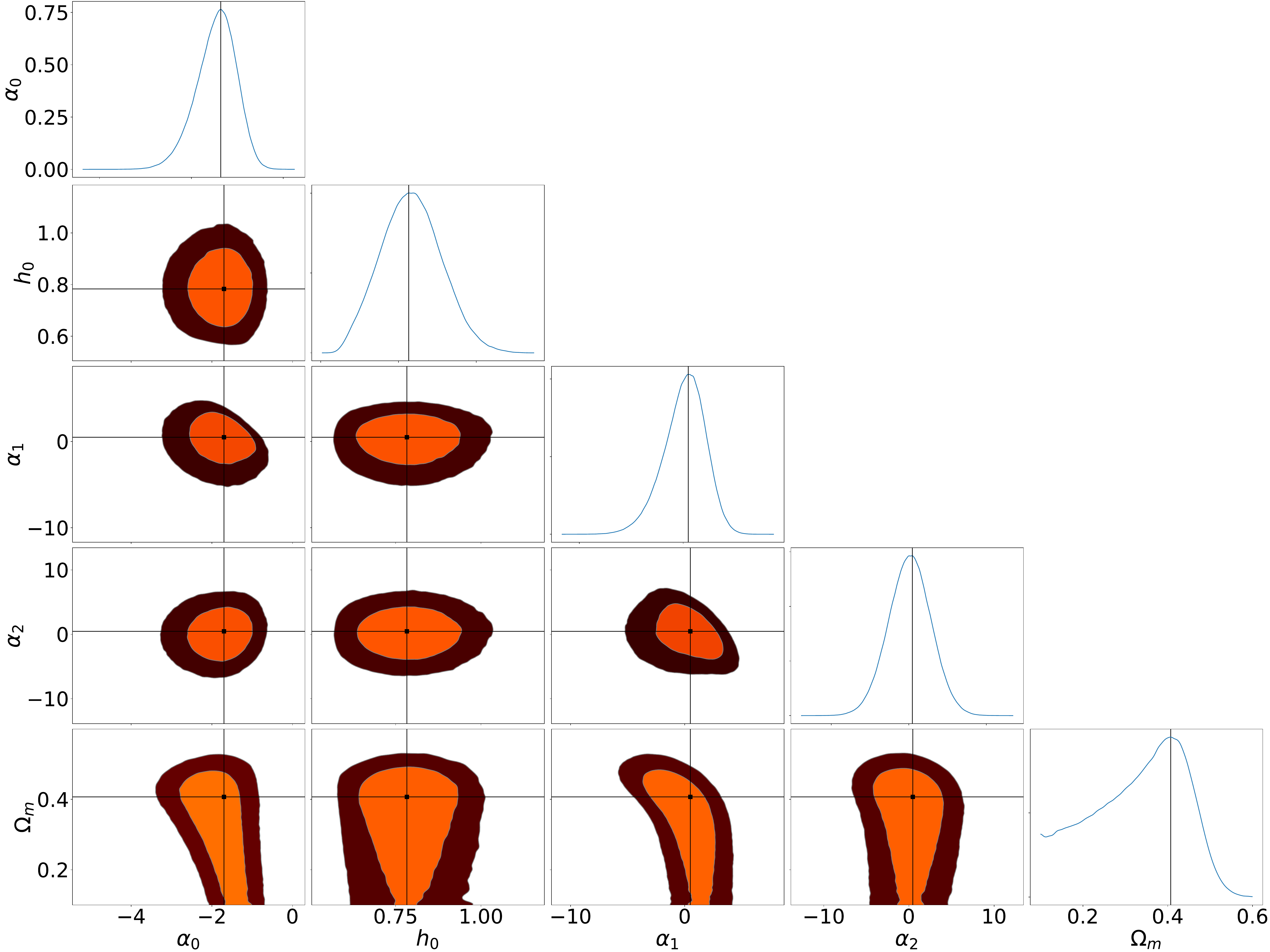}
    \caption{The plot shows the $1 \sigma$ and $2 \sigma$ contours for all the parameters along with their marginal probability density plots for the case of real data-set \citep{ohd1, ohd2, ohd3, ohd4, ohd5}. As in the previous figure, the first plot of every column is the marginal probability density plot.}
\label{fig:real_cont}
\end{figure*}

\begin{figure*}
\begin{center}
\includegraphics[scale=0.25]{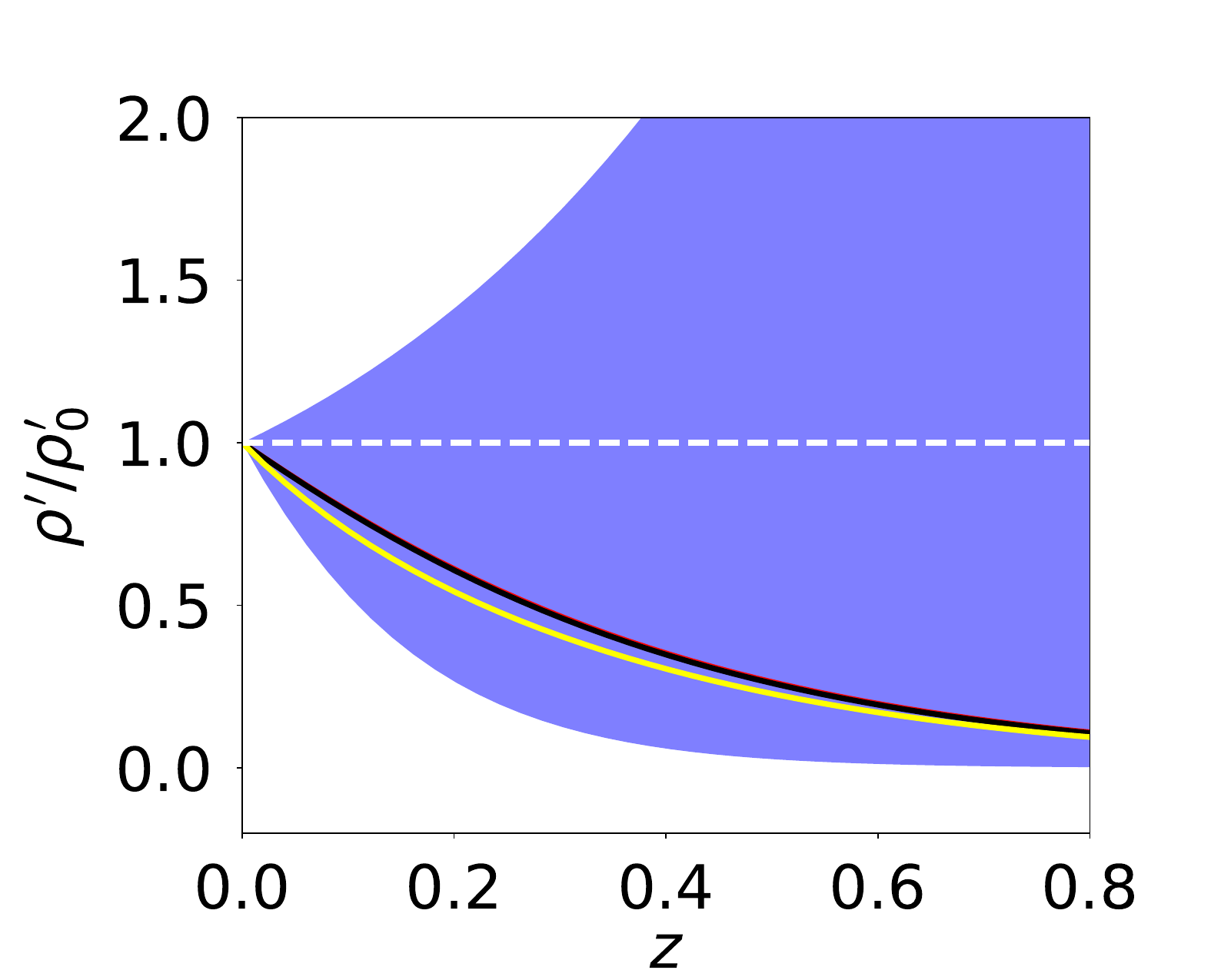}
\includegraphics[scale=0.25]{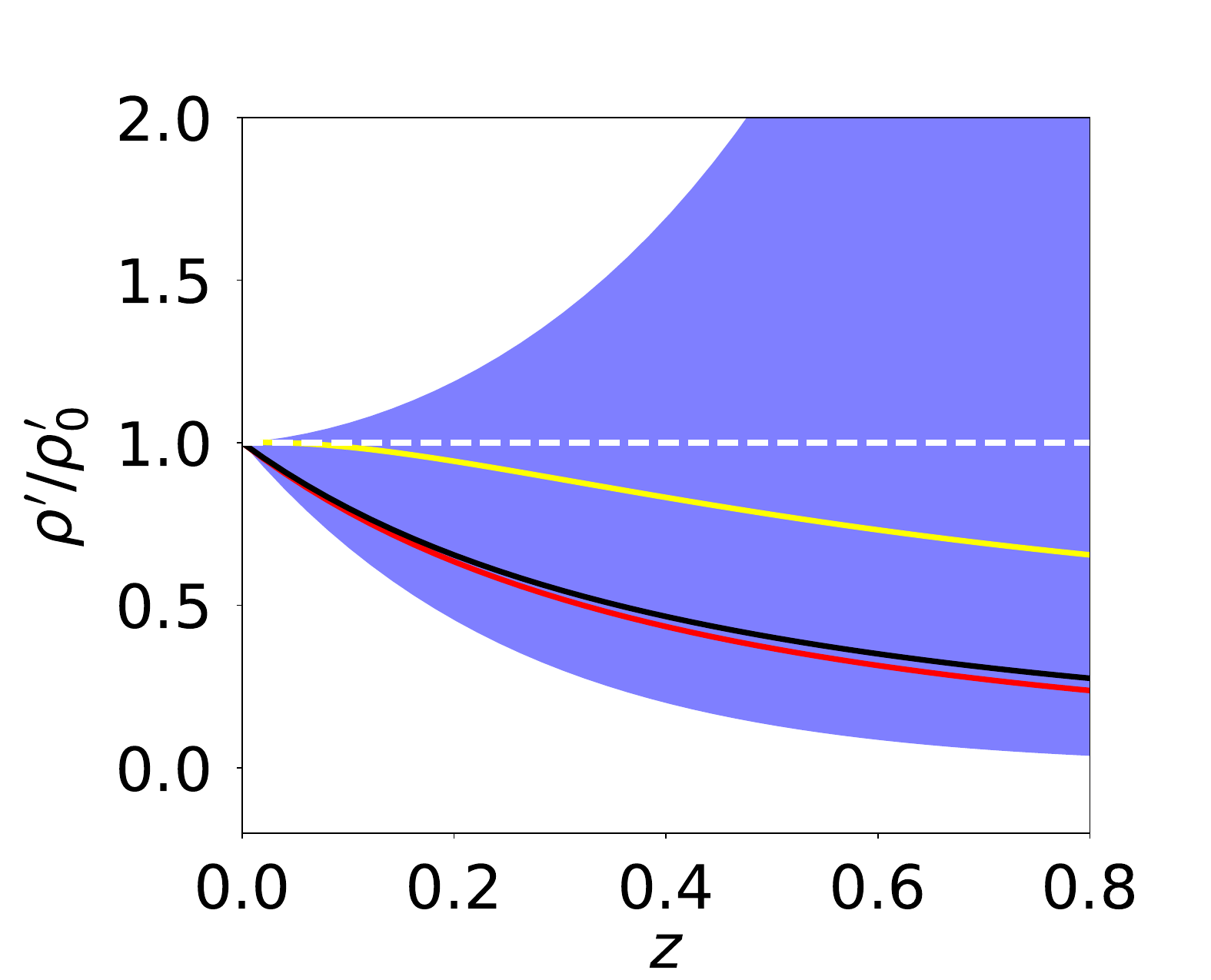}
\end{center}
\caption{In this figure, we show the  $68 \%$ confident level range for the dark energy density evolution $\frac{\rho^{\prime}}{\rho_0^{\prime}}$. Here, $\rho^{\prime}$ and $\rho_0^{\prime}$ are the dark energy density at redshift $z$ and at the present time, respectively.
Like in figure \ref{fig:EoS_ranges_simu_and_real}, the left figure is for the simulated Hubble parameter data set, and the right is for the real data-set.
The value of observational points $n_d$ and sample points $M_s$ are 600 and 800000, respectively.
The black, red, and yellow curves are the median, mean, and mode of the posterior density function.
The cosmological constant model is consistent with the data and is denoted by the dashed white line. 
For both the plots, mean and the median line overlap.}
\label{fig:DE_evolution_ranges_simu_and_real}
\end{figure*}

\begin{figure*}
	
\begin{subfigure}{0.3\textwidth}
		\centering
		\includegraphics[width=\linewidth] {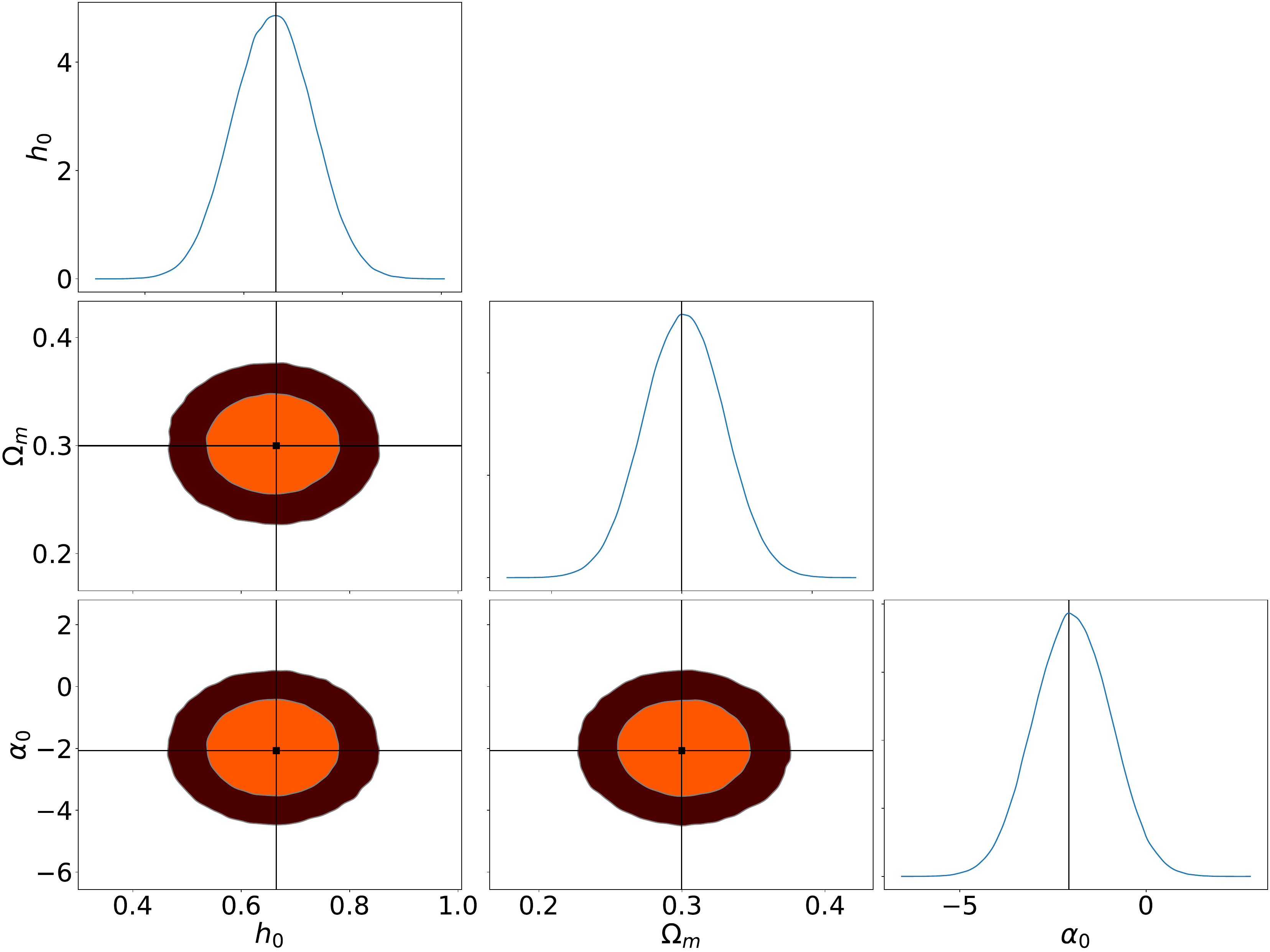}
\end{subfigure} 
\hfill 
\begin{subfigure}{0.6\textwidth}
	\centering 
		\includegraphics[width=\linewidth]{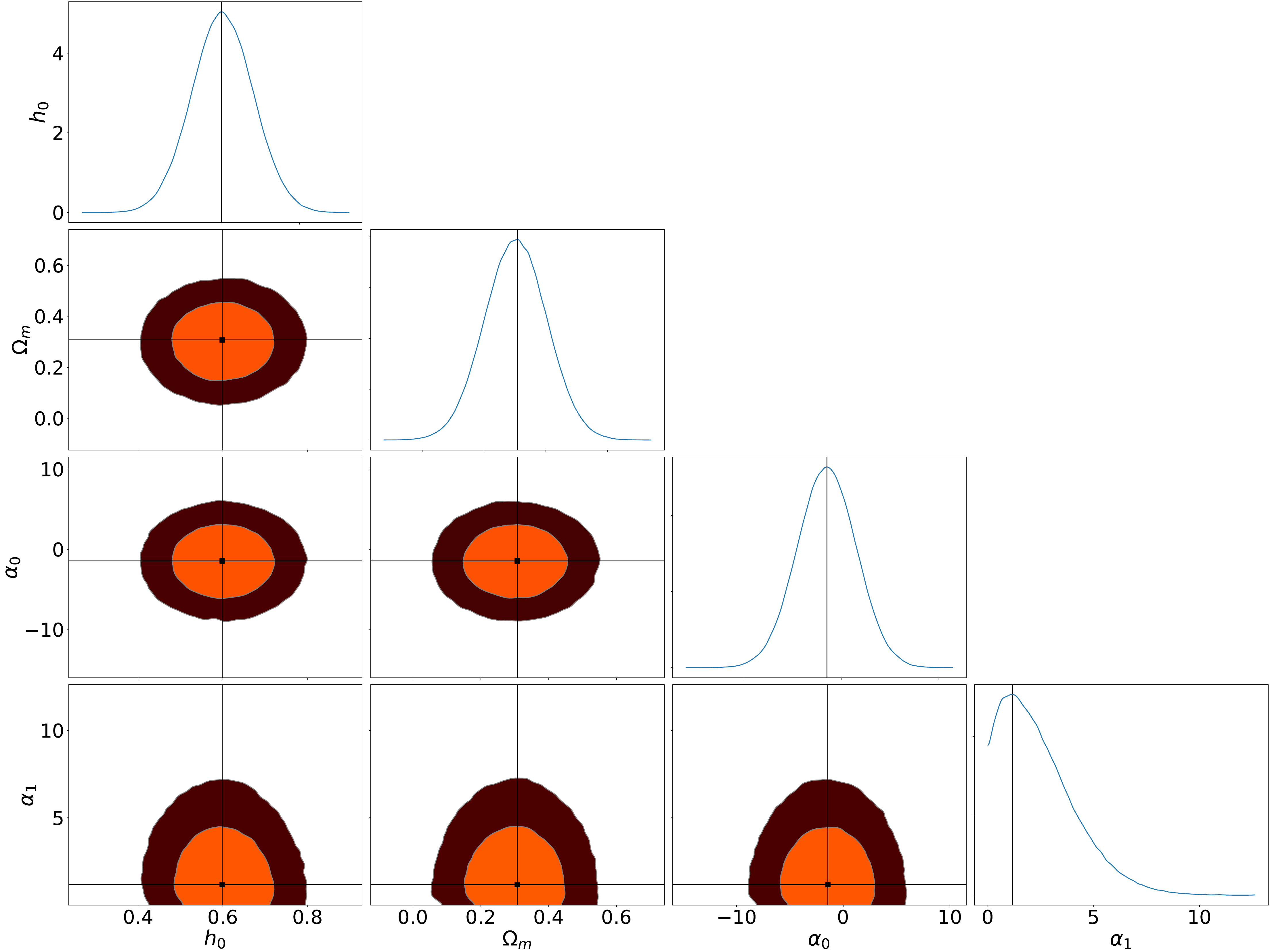}
\end{subfigure}

\caption{In this figure we show the $1 \sigma$ and $2 \sigma$ contours for all the parameters for the $w_{CDM}$(left) as well as for the $w_{CPL}$ model. 
Plots in the top and right of the main figures are for the marginal probability density plot for the SNIa data-set.
The plots are created for $[n_d, M_s] = [100, 80000]$. 
Here, we use the Cepheid Calibrated SNIa data-set\citep{Scolnic:2021amr, CC0_Riess_2021, Uddin_2023, CC1}.}
\label{fig:w_cpl & w_constant}
\end{figure*}

\begin{figure*}
		\includegraphics[width=\linewidth]{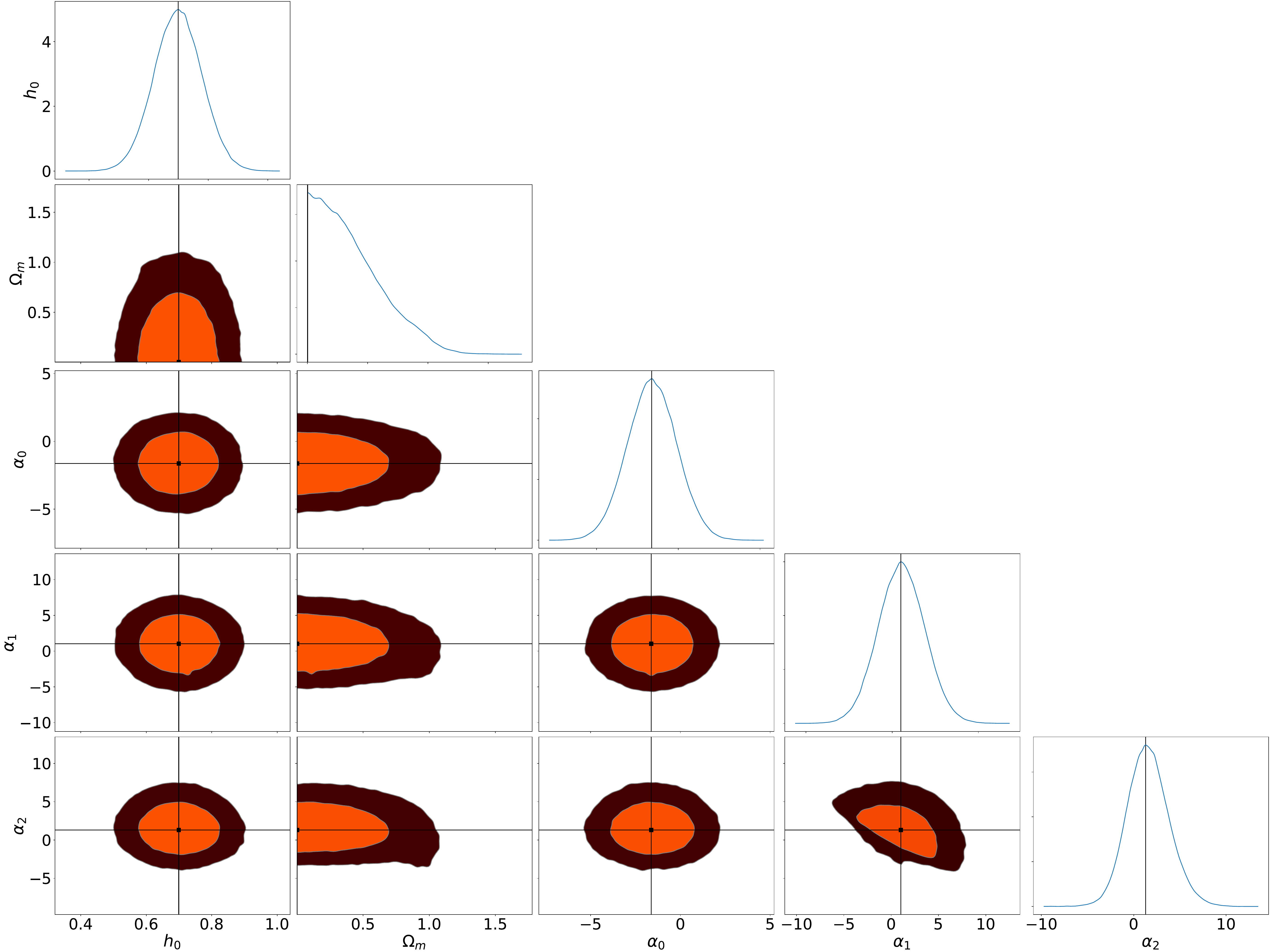}

\caption{The plot shows the $1 \sigma$ and $2 \sigma$ contours for all the parameters along with their marginal probability density plots for the case of distance modulus data-set \citep{Scolnic:2021amr, CC0_Riess_2021, Uddin_2023, CC1}. For the plot we consider $[n_d, M_s] = [100, 80000]$. The models considered here is the model described in eqn(\ref{eq:FRW}, \ref{eq:param_eos}).
Plots in the top and right of the main figures are for the marginal probability density plot for the SNIa data-set. }
\label{fig:panth_wConstant_and_wCPL}
\end{figure*}

\begin{figure*}
	\includegraphics[width=\linewidth, height=\textheight,keepaspectratio]{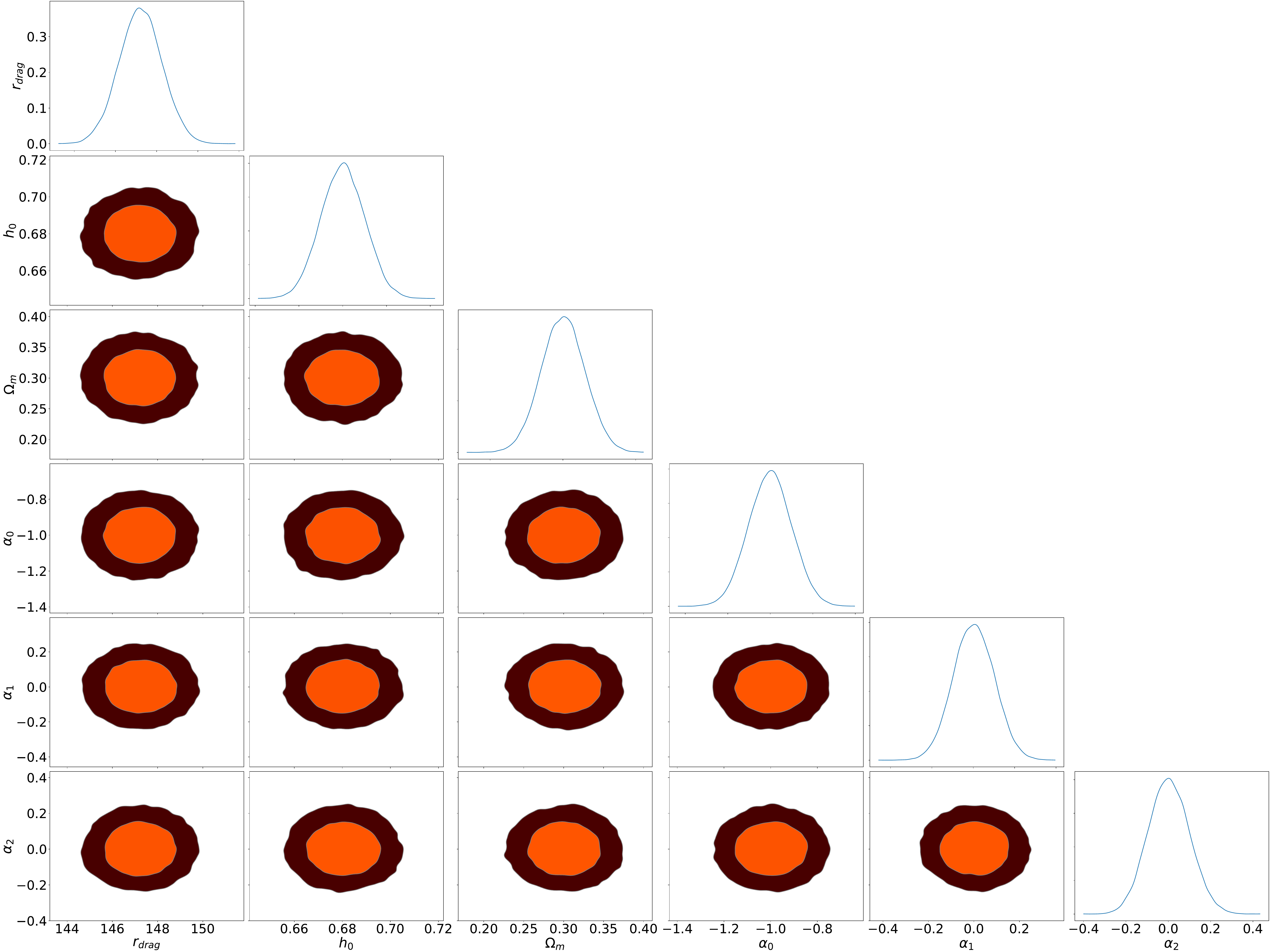}

\caption{The plot shows the $1 \sigma$ and $2 \sigma$ contours for all the cosmological parameters along with $r_{drag}$. Their marginal probability density plots for the case of transverse BAO data-set \citep{Scolnic:2021amr, CC0_Riess_2021, Uddin_2023, CC1} are shown in the first subplot of each column. For the plot we consider $[n_d, M_s] = [50, 10000]$. The models considered here is the model described in eqn(\ref{eq:FRW}, \ref{eq:param_eos}).}
\label{fig:bao_alone}
\end{figure*}

\begin{figure*}
		\includegraphics[width=\linewidth]{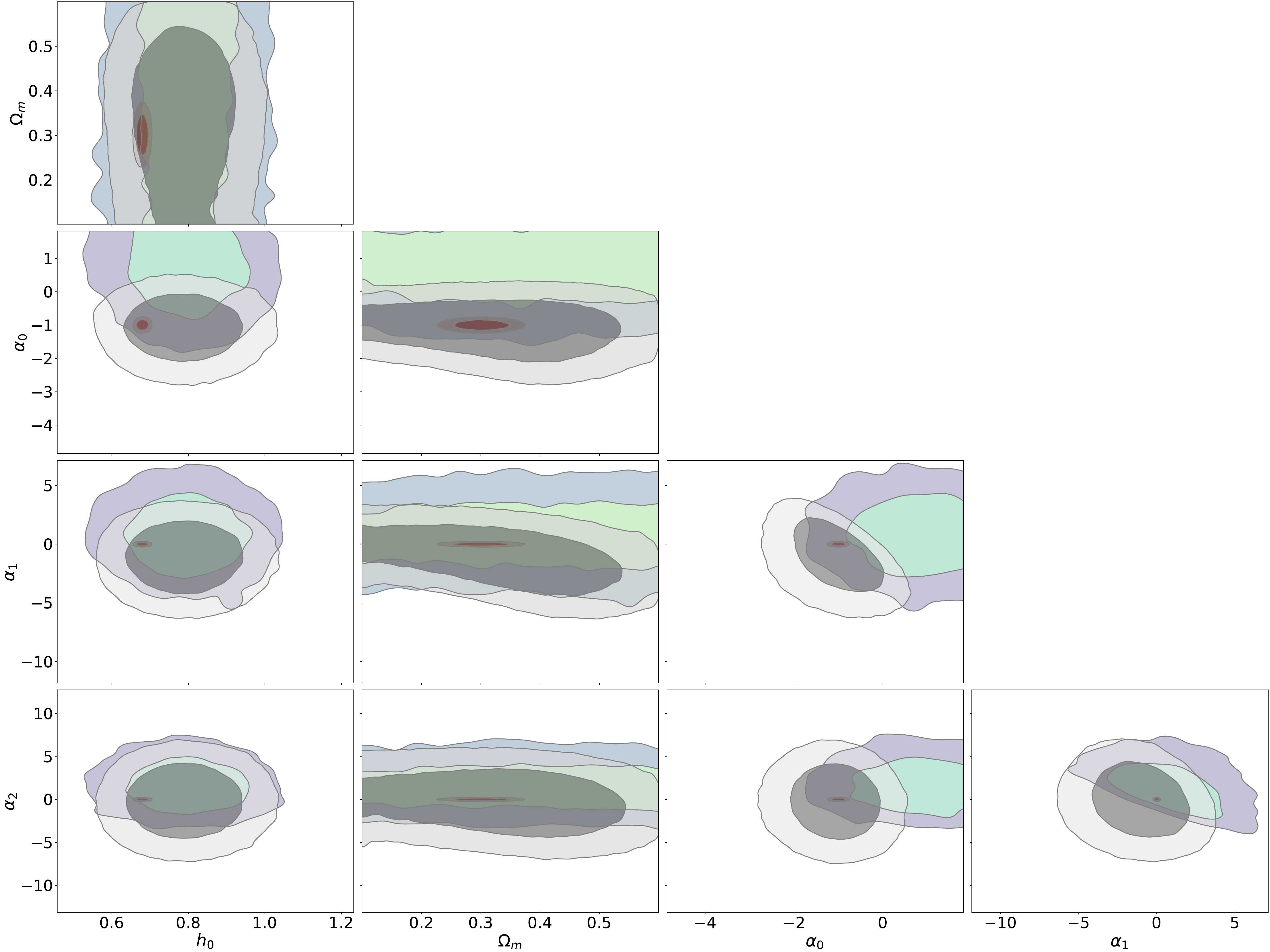}

\caption{The plot shows the joint likelihood assumed by all the data-sets. 
The $1 \sigma$ and $2 \sigma$ contours for Hubble parameters are shown in green-blue colors respectively, while the same for the SNIa are in grey colors, and for the BAO it is in  red. 
The model considered here is the model described in eqn(\ref{eq:FRW}, \ref{eq:param_eos}).
Here, for all the data-set we run the chains for $(n_d, M_s)=(50, 10000)$}.
\label{fig:comb_all}
\end{figure*}

\subsection{No-U-Turn sampler} 
To implement the MCMC search, we use the No-U-Turn sampler (NUTS), which effectively chooses the best parameter region. 
The No-U-Turn sampler is a modification of the Hamiltonian Monte Carlo(HMC), where the algorithm intrinsically selects the Leapfrog steps \citep{pymc, gelman-rubin-1992, Hoffman_Gelman}.
The selection of leapfrog steps is crucial in solving the Hamiltonian differential equations of the HMC.  
At every step, NUTS proceeds by creating a binary tree. 
In this binary tree, two particles representing progress in the forward and backward directions are created. 
If these two are represented as $(\mathbf{q_n^+}, \mathbf{p_n^+})$ and $(\mathbf{q_n^-}, \mathbf{p_n^-})$ then the NUTS conditions can be given by, 

\begin{align*}
(\mathbf{q}_n^+ - \mathbf{q}_n^-).\mathbf{p}_n^- < 0 \\
(\mathbf{q}_n^+ - \mathbf{q}_n^-).\mathbf{p}_n^+ < 0 
\end{align*}

In HMC, we move in the phase space of $\mathbf{q}$ and $\mathbf{p}$ in the elliptical path \citep{gelman-rubin-1992, Hoffman_Gelman}.  
The motivation for introducing the momentum variable $\mathbf{p}$ is to ensure we are exploring the greater area of the parameter space. 
This is done by moving in an elliptical contour, which we get after solving the dynamical Hamiltonian equation. 
In NUTS, when we move half of the elliptical path, the sign of the momentum and the position variables are changed, and we stop.
This makes the NUTS more efficient than HMC, wherein there is no way to ascertain if we are moving in the region of parameter space that is already explored.

We choose the value of the total sample points $M_s$ by checking the convergence limit using Gelman-Rubin statistic  \citep{pymc}.
Gelman-Rubin statistic for convergence is based on the notion that multiple convergence chain appears
to be similar to each other; otherwise, they will not converge.
It is a standard method to run multiple MCMC chains to test for convergence.
Scale reduction factor $\hat{r_o}$ is used to check the Gelman-Rubin convergence.  
There are two main ways the sequences of MCMC iterations fail to converge. 
In one case, the chains run in different parts, which have drastic differences in posterior probability densities of the target distribution. On the other, the chains fail to attain convergence.  
We change the value of $M_s$ until we get $\hat{r_o} = 1$, which is a confirmation of attaining the convergence. 

\section{Results} \label{sec_3}

We do the analysis described above for the Hubble parameter data, Cepheid Calibrated SNIa data and for the BAO data-set.
For the Hubble parameter, we show the results of both the simulated and the real data-set.
The simulated data-set is created using the same parameter values as is fixed by \citep{Aghanim:2018eyx}.  
For the simulated $\Lambda$CDM data-set we have fixed the values of cosmological parameters as $\Omega_m = 0.3$ and $h_0 = 0.685$.
We test the validity of our method and check if the analysis picks up these values. 
We then apply the method to the real data-set, namely  the Cosmic-Chronometer data-set \citep{MM0,MM1,MM2,ohd2,ohd3,ohd4,ohd5}
as well as SNIa data-set \citep{Scolnic:2021amr, CC0_Riess_2021, CC1, Uddin_2023}, then, compare with the usual likelihood analysis results.
We also use the transverse BAO data-set from  \cite{bao1,bao2,bao3,bao4}, which consists of  15 transverse BAO measurements \citep{bao101}, that are calculated using the public data-releases of the Sloan Digital Sky Survey (SDSS)\citep{sdss_2000}, without assuming a fiducial cosmological model \citep{bao_trans1,bao_trans2}.

To get the reconstructed curve of reduced Hubble parameter, distance modulus, as well as angular scale
we use $f(z) = \frac{z}{(1 + z)}$ as basis variable for simulated as well as 
observational dataset.
This initial basis function gives the best reconstruction as shown  in  \citep{rah_2020}.
Here, we have the freedom to choose the value of $n_d$,
which is the number of data points in the observed part of MLE.
We run the Markov Chain Monte Carlo (MCMC) chain to search for minimum $\chi^2$, which gives us the likelihood of the PCA data-set. 
In the MCMC analysis, for Hubble parameter data-set we use normal 
priors $\mathcal{N}(0.70, 0.2)$ and $\mathcal{N}(0.35, 0.1)$
for reduced Hubble constant $h_0$ and $\Omega_m$ respectively. 
For the DE parameters, $\vec{\alpha}$ we take $\mathcal{N}(0, 3)$. 
Here, $\mathcal{N}(x_{mean}, x_{mode})$ represents the normal probability density function with mean $x_{mean}$ and spread of 
$x_{mode}$.
For Cepheid Calibrated SNIa data, we use the data archive given in \cite{CC0_Riess_2021, Uddin_2023, CC1}.
In the MLE part, we take half normal with a standard deviation of $0.4$ as a 
prior of $\Omega_m$ and for $\Vec{\alpha}$ we take $\mathcal{N}(-2, 1.5)$.
In the case of the BAO data-set, we use the same priors as for Hubble parameter dataset.

\begin{table*}
\centering
  \begin{tabular}{ |m{3cm}|m{2.5cm}|m{2.5cm}|m{2.5cm}|m{2.5cm}|}
    \hline
data-type and model & parameter & \hspace{5mm}  1$\sigma$ & \hspace{5mm}  2$\sigma$ & \hspace{5mm} best-fit \\

\hline
\multirow{2}{*}{Hz Simu ($w_{model}$)} & \hspace{5mm} $h_0$ & [0.55, 0.7161] & [0.496, 0.847]  & 0.674 \\
\cline{2-5}
& \hspace{5mm} $\Omega_m$ & [0.238, 0.415] & [0.169, 0.508] & 0.328 \\
\cline{2-5}
& \hspace{5mm}  $\alpha_0$ & [-3.01, -0.48] &[-4.24, 0.739] & -1.74\\
\cline{2-5}
& \hspace{5mm}  $\alpha_1$ & [-4.379, 1.145] &[-7.047, 3.85] & 1.602\\
\cline{2-5}
& \hspace{5mm}  $\alpha_2$ & [-4.501, 1.394] &[-7.314, 4.3] & -1.54\\

\hline
\multirow{2}{*}{Hz Real ($w_{model}$)} & \hspace{5mm} $h_0$ & [0.689, 0.885] & [0.601, 0.974]  & 0.784 \\
\cline{2-5}
& \hspace{5mm} $\Omega_m$ & [0.247, 0.473] & [0.107, 0.488] & 0.328 \\
\cline{2-5}
& \hspace{5mm}  $\alpha_0$ & [-2.278, -0.998] &[-2.94, -0.83] & -1.84\\
\cline{2-5}
& \hspace{5mm}  $\alpha_1$ & [-4.492, 1.023] &[-7.143, 3.75] & 0.08\\
\cline{2-5}
& \hspace{5mm}  $\alpha_2$ & [-1.59, 2.296] &[-5.33, 5.422] & 0.09\\

\hline
\multirow{2}{*}{SNIa ($w_{CDM}$)} & \hspace{5mm} $h_0$ & [0.579, 0.739] & [0.503, 0.815]  & 0.703 \\
\cline{2-5}
& \hspace{5mm} $\Omega_m$ & [0.272, 0.332] & [0.242, 0.361] & 0.345 \\
\cline{2-5}
& \hspace{5mm}  $\alpha_0$ & [-3.01, -0.965] &[-3.98, -0.019] & -1.89\\

\hline
\multirow{2}{*}{SNIa ($w_{CPL}$)} & \hspace{5mm} $h_0$ & [0.52, 0.68] & [0.45, 0.76] & 0.645 \\
\cline{2-5}
& \hspace{5mm} $\Omega_m$ & [0.201, 0.401] & [0.105, 0.498] & 0.301 \\
\cline{2-5}
& \hspace{5mm}  $\alpha_0$ & [-4.49, -0.1.51] &[-7.40, 4.36] & -1.47 \\
\cline{2-5}
& \hspace{5mm}  $\alpha_1$ & [0.0072, 3.02] &[$\mathcal{O}(-5)$ , 5.78] & 2.44\\
\cline{2-5}

\hline

\multirow{2}{*}{SNIa ($w_{model}$)} & \hspace{5mm} $h_0$ & [0.618, 0.777] & [0.542, 0.855]  & 0.696\\
\cline{2-5}
& \hspace{5mm} $\Omega_m$ & [$\mathcal{O}(-5)$, 0.46] & [$\mathcal{O}(-5)$, 0.894] & 0.329 \\
\cline{2-5}
& \hspace{5mm}  $\alpha_0$ & [-3.12, -0.122] & [-4.56, 1.36] & -1.58 \\
\cline{2-5}
& \hspace{5mm}  $\alpha_1$ & [-1.65, 3.7] &[-4.24, 6.33] & 0.957\\
\cline{2-5}
& \hspace{5mm}  $\alpha_2$ & [-0.809, 3.62] &[-2.81, 6.174] & 1.43\\
\cline{2-5}

\hline

\multirow{2}{*}{BAO ($w_{model}$)} & \hspace{5mm} $h_0$ & [0.67, 0.69] & [0.66, 0.699]  & 0.68\\
\cline{2-5}
& \hspace{5mm} $\Omega_m$ & [0.269, 0.328] & [0.241, 0.351] & 0.30 \\
\cline{2-5}
& \hspace{5mm}  $\alpha_0$ & [-1.10, -0.903] & [-1.2, -0.81] & -1.0 \\
\cline{2-5}
& \hspace{5mm}  $\alpha_1$ & [-0.092, 0.104] &[-0.192, 0.19] & 0.0\\
\cline{2-5}
& \hspace{5mm}  $\alpha_2$ & [-0.099, 0.096] &[-0.19, 0.196] & 0.01\\
\cline{2-5}

\hline

\end{tabular}

\caption{This table gives the 1$\sigma$ and 2$\sigma$ ranges for parameters, $h_0$, $\Omega_m$ and $\vec{\alpha}$, for all the different model and the data-type for which we run our complete analysis.  
For the cosmic chronometer data-set we use both the simulated and the real data-sets and apply it to $w_{model}$. 
For SNIa data we do our analysis for 
 $w_{CDM}$, $w_{CPL}$ as well as the dark energy model of eqn(\ref{eq:param_eos}) with $m=3$. Here $\vec{\alpha}$ are the parameters of dark energy equation of state parameter.
The last column of the table corresponds to the best-fit value of the parameter of the model given the data-set.}

\label{Table::Tot}

\end{table*}



We choose the largest possible value for $n_d$, which is limited by the computing power. 
We then check the results for different values of $n_d$ and $M_s$.
Moreover, we find out the posterior distribution's mean, median, and mode.
For $m = 3$ in eqn(\ref{eq:param_eos}), we do the analysis for different values of $n_d$ and $M_s$.
$m=3$ is the CPL parameterization along with the next order term \citep{cpl0, cpl1}.
We vary $n_d$ in the range 100 to 800 whereas $M_s$ in the range 1000 to 800000 and find out 
mean, median, and mode as well as 1$\sigma$ and 2$\sigma$ ranges of $\omega_m$, $h_0$, $\vec{\alpha}$.

From the mode plot of the posterior of the model likelihood of fig(\ref{fig:EoS_ranges_simu_and_real}) and fig(\ref{fig:DE_evolution_ranges_simu_and_real}), we can see the difference between  the old cosmic chronometer data-set (\cite{ohd1,ohd2,ohd3,ohd4,ohd5}) with the new cosmic chronometer data-set(\cite{MM0,MM1,MM2}). 
For the particular cosmological model of eqn(\ref{eq:param_eos}), with the NUTS algorithm, PCA reconstruction brings $w(z)$ closer to the $w(z) = -1$ in comparison to the old cosmic chronometer data-set. 

In Figures \ref{fig:real_cont} and \ref{fig:simu_cont}, we show results for $n_d = 600$, where we fix
the number of sample points at $M_s = 80,0000$.
This particular choice of $n_d$ and $M_s$ gives us the closest approximation
of the model parameters for the simulated Hubble parameter data. 
Also, we see that about this value of $n_d$ and $M_s$ we get the smallest variation in $1 \sigma$ and $2 \sigma$ ranges of the model parameters, with the variation of these two quantities.
In particular for $(n_d, M_s) = (600, 800000) \& (1000, 50000)$ the difference in $1 \sigma$ and
$2 \sigma$ ranges are of the order of $\mathcal{O}(-1)$ for $\vec{\alpha}$ and  
$\mathcal{O}(-2)$ or less for $\Omega_m$ and $h_0$.  
For the Hubble parameter data-set, 
the mean of the posterior of 
$h_0$ and $\Omega_m$ from the algorithm are $h_0 = 0.68$ and $\Omega_m = 0.34$, which are very close to the 
assumed values to produce the simulated data-set, $h_0 = 0.685$ and $\Omega_m = 0.3$.
In table \ref{Table::Tot} we show the $1\sigma$ and  $2\sigma$ ranges for the parameters, along with their best-fit values.
The mean of the posterior of $h_0$ and $\Omega_m$ for the real Hubble parameter data  are $h_0 = 0.71$ and $\Omega_m = 0.35$, respectively.
Also, we present our results for SNIa data-set \citep{Scolnic:2021amr, CC0_Riess_2021, Uddin_2023, CC1} as well as BAO data-set \citep{bao1,bao2,bao3,bao4,bao101}. 
For SNIa data-set, we show our results for $w_{CDM}$ and $w_{CPL}$ in the fig(\ref{fig:w_cpl & w_constant}).
For the model of eqn(\ref{eq:param_eos}) with $m=3$, we show our results in fig(\ref{fig:panth_wConstant_and_wCPL}).
We present our results for $(n_d, M_S) = (100, 80000)$ for the SNIa data-set.
Table(\ref{Table::Tot})  gives a  comparison; this table can be extended to different data-sets and models.

In fig(\ref{fig:bao_alone}), we present results for the BAO data-set, using the DE model with $m=3$ of eqn(\ref{eq:param_eos}).
We see, both from the fig(\ref{fig:bao_alone}), and the combined joint analysis plot fig(\ref{fig:comb_all}) that BAO gives tighter constraints in comparison to Hubble parameter and SNIa data-sets. 
For  fig(\ref{fig:comb_all}) and fig(\ref{fig:bao_alone}) we use $(n_d, M_s)=(50, 10000)$, and this choice is done under the available computational power.
From the fig(\ref{fig:bao_alone}) we can see that PCA + MCMC  gives very good reconstruction even with a  small number of sample points $M_s$.
The 1$\sigma$ range of $r_{drag}$ from our method is $[146.1, 148.2]$. 

It is also evident from the fig(\ref{fig:EoS_ranges_simu_and_real}, \ref{fig:real_cont}, \ref{fig:DE_evolution_ranges_simu_and_real}, \ref{fig:w_cpl & w_constant}, \ref{fig:panth_wConstant_and_wCPL}, \ref{fig:bao_alone}) that $w(z) = -1 $ is well within the $1\sigma$ 
range of $w(z)$ parameters($\vec{\alpha}$).
The plot of $w(z)$ and $\rho(z)/\rho_0$ are similar for both real and simulated data-set.
The difference in $w(z)$ and $\rho(z) / \rho_0$ curve between simulated and real Hubble parameter data
are 0.445 and 0.026, respectively.
Here, $\rho(z)$ and $\rho_0$ are the total energy density at redshift $z$ and at present. 
The dark energy density plot, $\rho^\prime(z) / \rho^\prime_0$
vs. $z$, for simulated and real Hubble parameter data-set are also similar, and the maximum difference between them for Hubble parameter data-set is $0.31$. 

We restrict to the $m=3$ cut-off in eqn(\ref{eq:param_eos}) for $(n_d, M_s)=(600, 800000)$, largely due to the computational power available to us at present. 
In the table \ref{Table::Time_cpu}, for $(n_d, M_s)=(100, 100)$ we use the algorithm upto $m=10$.
We find out that for the Hubble parameter data-set the better constraint on the parameter space for $m \geq 4$ needs to be done with $(n_d, M_s) \geq (600, 800000)$.  
In a follow-up work we are trying to optimize the algorithm to constrain the parameter space with large enough values of $m$, and draw the physical conclusion over it.
The parametrization of $w(z)$ and $\frac{\rho^{\prime}_{de}}{\rho^{\prime}_{0}}$ have the same physical implication, which is shown in fig(\ref{fig:EoS_ranges_simu_and_real}, \ref{fig:DE_evolution_ranges_simu_and_real}).
The dark energy density can be derived analytically for these parameterizations, and fixing the dark energy equation of state parameter determines the evolution of the energy density as a function of time.

In the MCMC run for both the real and simulated Hubble parameter data-set, with $(n_d, M_s) = (600, 800000)$, 
the value of Gelman-Rubin convergence factor $\hat{r_o}$ is $1$.
For SNIa MCMC run $(n_d, M_s) = (100, 80000)$ gives the value of $\hat{r_o} = 1$. 
To check the convergence, we not only check the $\hat{r_o}$ factor and eliminate those iterations 
which do not satisfy the $\hat{r_o} \approx 1$ criteria, but we also check the trace plots, rank bar 
plots and the rank vertical line plots of the posterior sampling for visual confirmations 
\citep{gelman-rubin-1992, brooks-gelman-1998,cowles-carlin-1996}. 

The error bars of the parameters, in table (\ref{Table::Tot}), are derived 
when error function from PCA is considered. 
Hence the  $1\sigma$, $2\sigma$ ranges are affected by the error functions we introduce in the MLE. 
For $w_{cpl}$, with the Pantheon data-set, when we consider a half-Normal probability distribution for the error part of
the MLE, the range of $h_0$ changes to $[0.6005, 0.6584]$, which is almost three times smaller than the range when PCA error function is considered. 
PCA error function is created solely from the data structure we provide in the first step of PCA. 
With the improvement of error-bars in the original data-points the range of the parameters will reduce significantly.



We also analyze with the classical Metropolis-Hastings (MH) as well as the Hamiltonian Monte Carlo sampler (HMC).  
For comparison with MH and HMC, we do the analysis with $(n_d, M_s) = (600, 800000)$ and $(n_d, M_s) = (100, 80000)$ for Hubble parameter and Supernovae data respectively.
Our analysis shows that NUTS and Hamiltonian Monte Carlo sampler (HMC) perform better than Metropolis-Hastings sampler (MH).
For more than six continuous parameters and with the same CPU power, NUTS improves speed by a factor of 2.4 to complete the analysis. 
Details of the time taken by the MH and NUTS are  given in  \ref{MH_plots}
The plots for the real Hubble parameter data-set, with the MH is shown in \ref{MH_plots}
Again, NUTS is better than HMC as after picking up the leapfrog steps, NUTS stops automatically when the NUTS conditions are satisfied.
It has been explicitly shown in \citep{Hoffman_Gelman} that the NUTS algorithm is more efficient. Convergence plots for the NUTS sampler in the case of the Hubble parameter data-set are shown in the \ref{covergence_plots} and that of the MH is shown in \ref{MH_plots}. 

\section {A comparison with other methods} \label{sec_new}
The reconstruction of $H(z)$, $\mu(z)$ and $\theta_b(z)$ from the PCA algorithm, which is described in the sec(\ref{sec_2}) and in \cite{rah_2020} is qualitatively different from the other PCA techniques employed in the literature \cite{Huterer2003prl,clarkson2010prl,Huterer2005,Zheng:2017ulu,Ishida2011is}.
The starting assumption  is that the function $h(z)$, $\mu(z)$ and $\theta_b$ are smoothly varying over redshift $z$.
This is a reasonable choice as described by the current data-sets, \citep{MM0,MM1,MM2,ohd2,ohd3,ohd4,ohd5,Scolnic:2021amr, CC0_Riess_2021, CC1, Uddin_2023, bao1, bao2, bao3, bao4}. 
Different variants of PCA techniques have been adopted in \cite{Qin:2015eda} and \cite{Liu:2015yha}. 
Before creating a different set of simulated Hubble data to construct the covariance matrix \cite{Qin:2015eda}
uses an error model. 
While \cite{Liu:2015yha}  use the weighted least square method and combine it
with PCA. 

We apply MLE to the observed data-sets with the reconstruction functional form of PCA.
Including a Cosmological model is only in the final part of our methodology, where we use the MLE technique.
MCMC chain with the NUTS, over the model-independent reconstruction of PCA gives unbiased constraints on the 
model parameters. 
Fisher matrix computation is one of the major ways to PCA reconstruction \citep{Nesseris:2013PhRvD,Huterer2003prl,clarkson2010prl,Huterer2005,Zheng:2017ulu,Ishida2011is,Crittenden:2005wj,
Miranda:2018,Hart:2019,Hojjati:2012prd,Nair2013,Hart:2021kad}.  
Our methodology calculates the Covariance matrix, which quantifies the correlation and uncertainties directly from 
the PCA data matrix described in the section \ref{sec_2}.
Reduction of dimension, which is a distinctive feature of the PCA reconstruction, omits the noise part from the PCA data matrix. 
Therefore, the parameter constraints that are done by replacing the observational part with the PCA reconstructed part will 
constrain the parameter in a more reasonable way. 
One important feature of the (PCA + MCMC) methodology is that it will also work for sparse data sets.
In comparison to the classical techniques, it can be easily generalized to a higher dimension of parameter space with 
little expense of computational time, as described in sec(\ref{sec_2}) and showed quantitively in \ref{appendix}.

\section{Conclusions} \label{sec_4}
In this paper, we combine the Principal Component Analysis reconstruction with the Markov Chain Monte Carlo  tool  to determine cosmological parameters.
We assume the Taylor series expansion of the equation of state parameter in terms of the scale factor as 
the parameterization of the dark energy equation of state. When the method of  PCA, along with correlation coefficient calculation is combined with the MCMC tool, we have the freedom of selecting the number of points in the observational part of maximum  likelihood method.
We use the No-U-Turn Sampler for this analysis.

We first test the method on simulated data and check if the values assumed for the cosmological parameters are reconstructed effectively.
We see that the predictions for the model parameters are consistent with the assumed values. 
The parameter estimation does not depend strongly on the prior probability assumption, and the idea can be generalized to other data-sets as well as different sampling techniques. 
The relation between the Hubble parameter and the equation of state of dark energy also contains the first differentiation of the Hubble parameter, which introduces an unwanted error in the equation of state predictions. 
Similarly, for SNIa and BAO data-sets the relation between the distance modulus as well as the angular scale with the EoS of dark energy contains first and second order differentiation.

The present method eliminates the error that arises from the first and higher order differentiation of the observable to infer the value and ranges of the Equation of State of dark energy. 
In this work, we only use the error function that comes directly from the PCA algorithm, and one can use different error functions in the error part of the MLE as well. 
It is clear from the results that, for the simple model of dark energy we take, the allowed range of cosmological parameters is consistent with other analyses, and the cosmological constant model is well within the allowed range of models for both the Hubble parameter and distance modulus data-set.  
This analysis can be extended to other dark energy models.
Here the advantage is that the complete functional form of the observable of the dataset is obtained; in the present work, the Hubble parameter and distance modulus as a function of redshift are determined. 
The second step is deriving the dark energy equation of state parameter. 
The method is suitable for different types of data and merits future analysis. 
It depends only on the model-independent reconstruction of the data-set and the error associated with it. 
Improvement of even a single data-point leads to an increase of the constraining power of our method. 
With the upcoming improved Hubble parameter, SNIa and BAO data-sets, the application of the method will lead
to better constraints and a much easier distinction between different dark energy models.
Also, the method discussed here can be used as a model selection tool in those data-sets with fewer data-points.


\vspace{5em}

\appendix

\label{appendix}

\section{Checks for convergence}
\label{covergence_plots}

Here, we show the trace-plot, rank bar plots and vertical line plots for the NUTS in the case of both simulated and real Hubble parameter dasta-sets. 
These plots are essential to check the convergence of the sample as well as the efficiency of the sampler. 

\begin{figure*}
	\includegraphics[scale=0.50]{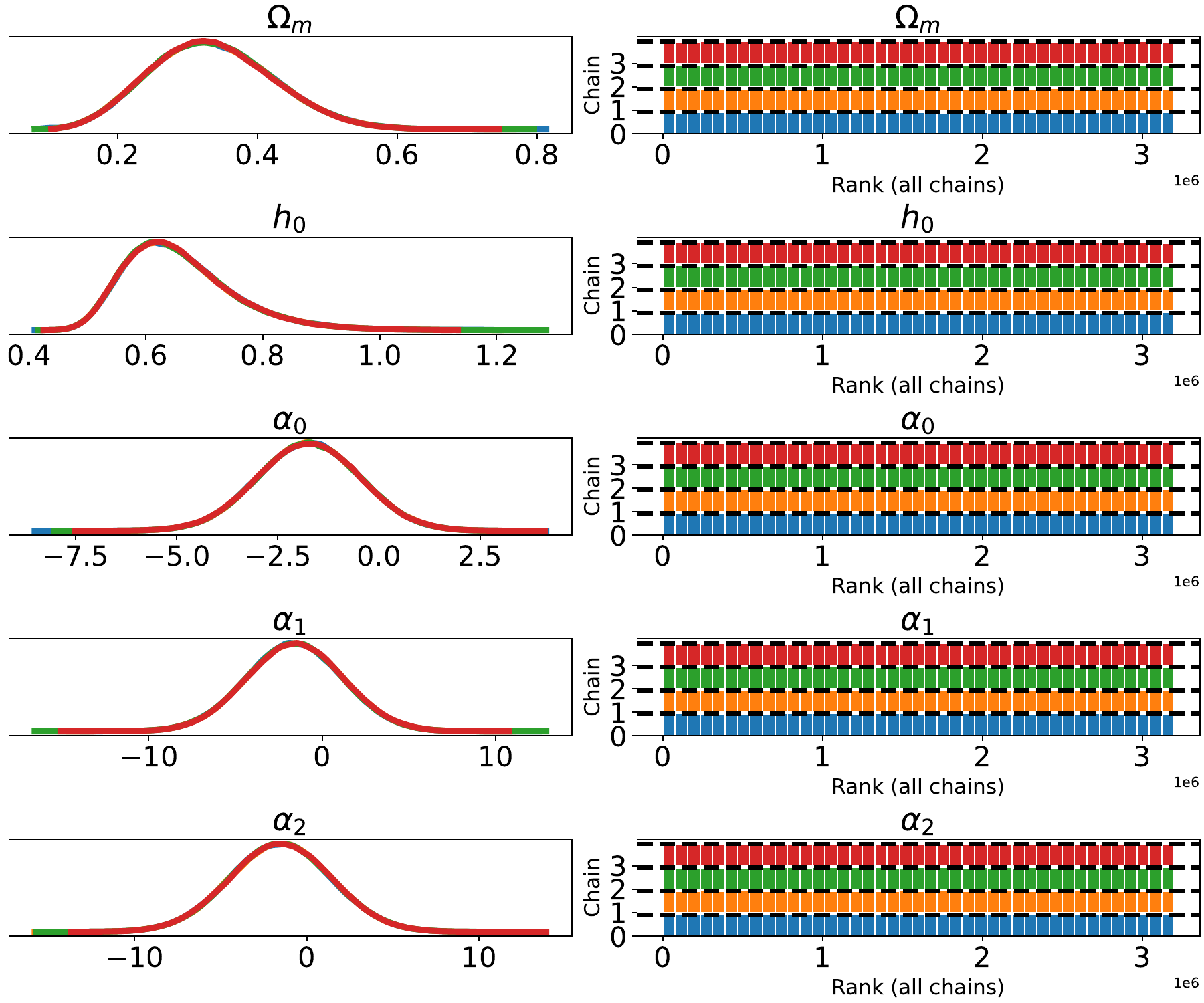}
	\caption{Smoothed histogram 
	using kernel density estimation of the marginal posteriors of each stochastic random variable. 
	From the top to bottom, the random variables are density parameters for matter $\Omega_m$, 
	reduced Hubble constant $h_0$, and the parameters of the equation of state of dark energy $\alpha_i$. 
	The right column shows the number of successful draws from the parameter space, which depends upon the sample size chosen as well as the prior theory. 
	The four MCMC chains are plotted in different colors and all are independent of each other.
	} 
\label{fig:600_800000_simu_rank_bar}
\end{figure*}

\begin{figure*}
	\includegraphics[scale=0.50]{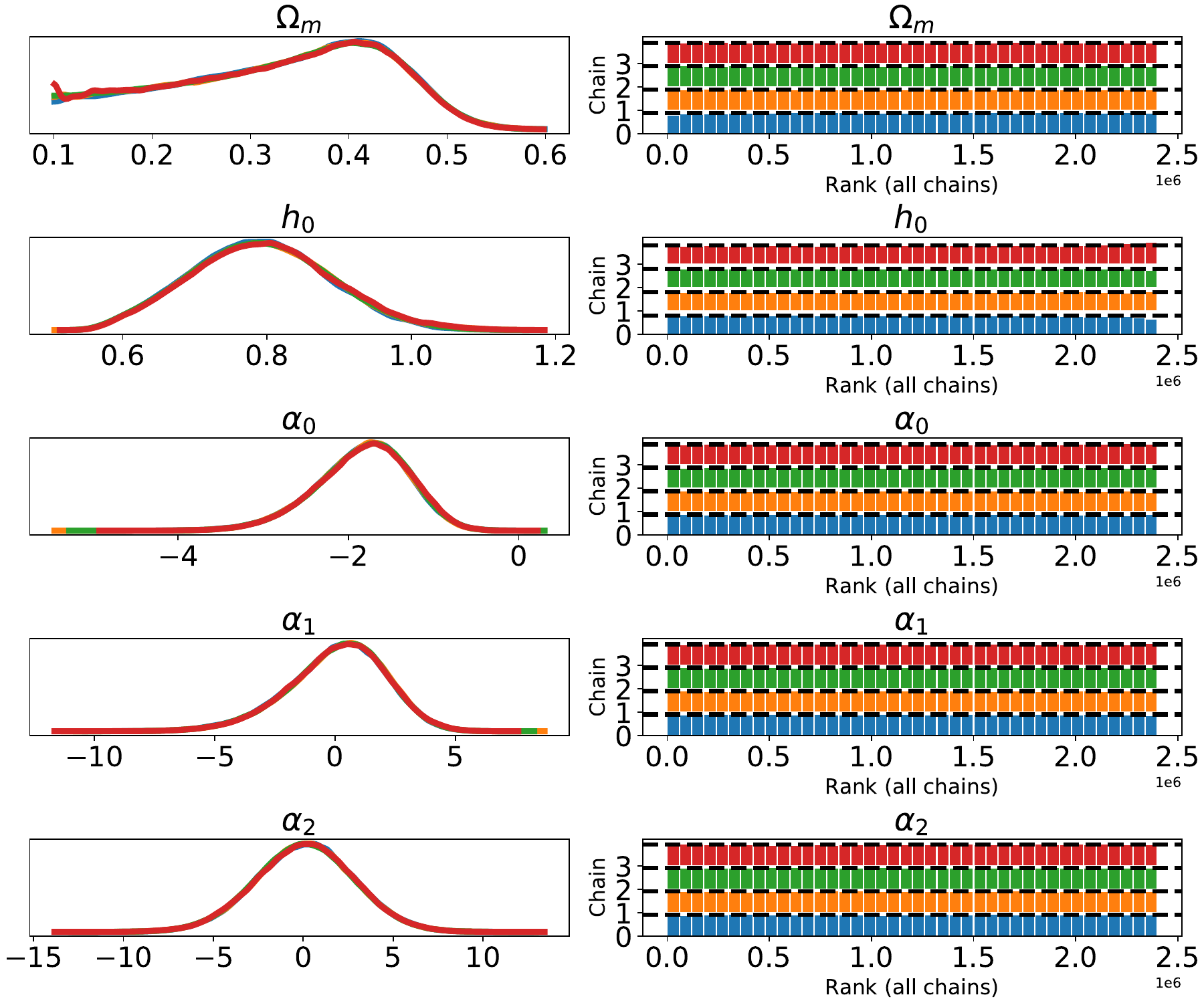}
	\caption{ The random variables plotted here are the same as in the previous figures and in the same order.
	The right column shows the number of successful draws from the parameter space, which depends upon the sample size chosen as well as the prior theory. 
	The four MCMC chains are plotted in different colors and all the chains are independent of each other.
	} 
\label{fig:600_800000_real_rank_bar}
\end{figure*}

As introduced in \cite{gelman2019}, rank plots are histograms of the rank posterior draws, plotted separately for each chain. 
In our case all the four chains are targeting the same posterior, hence we expect the ranks in each chain to be uniform.   
The similar rank plots of fig(\ref{fig:600_800000_simu_rank_bar}) and fig(\ref{fig:600_800000_real_rank_bar}) indicates good mixing of the chains.

\begin{figure*}
	\includegraphics[scale=0.50]{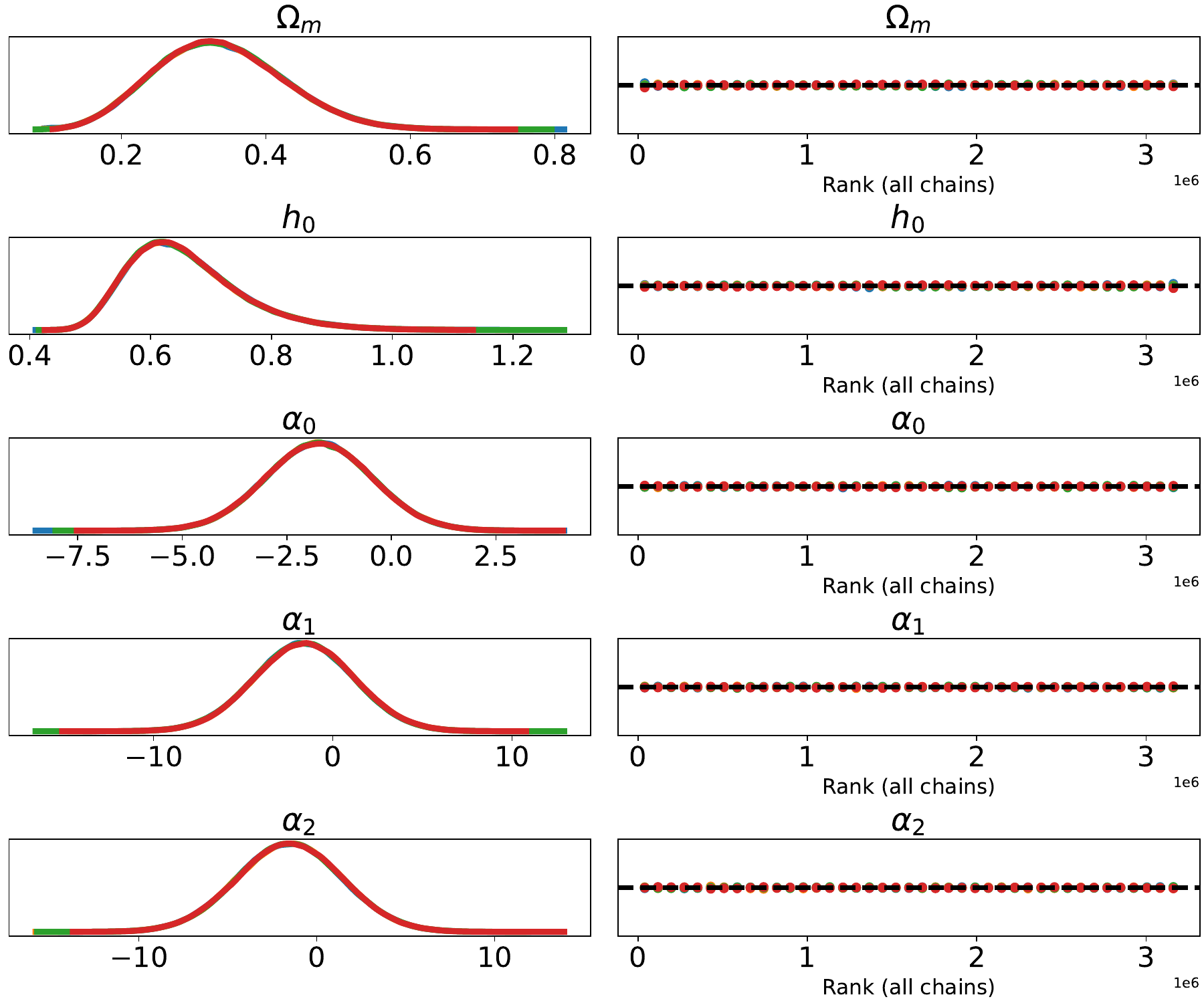}
	\caption{In this figure, the left column consists of a smoothed histogram, 
	using kernel density estimation of the marginal posteriors of each stochastic random variable. 
	The right-hand column gives the auto-correlation for all the random variables for all four independent chains, which are depicted with vertical lines.  
	} 
\label{fig:600_800000_simu_vlines}
\end{figure*}

To assess the MCMC methods we have to measure how good the MCMC estimates are, which could be done using autocorrelation time, the variance of the estimate, or the effective sample size. 
The autocorrelation of a variable measures the relationship between its present value with any of the past value which we can access. 
If instead of one value we compare the current series of values with the past historical data it is called autocorrelation time series. 
Fig(\ref{fig:600_800000_simu_vlines}) and fig(\ref{fig:600_800000_real_vlines}) gives the autocorrelation for all the four chains.
The classical sample techniques like Metropolis-Hasting or Hamiltonian Monte Carlo create autocorrelated samples if the number of continuous parameters are very large. 
NUTS works very effectively in the case of parameter space which consists of large number of continuous variables  \citep{pymc, gelman-rubin-1992, Hoffman_Gelman}.

\begin{figure*}
	\includegraphics[scale=0.50]{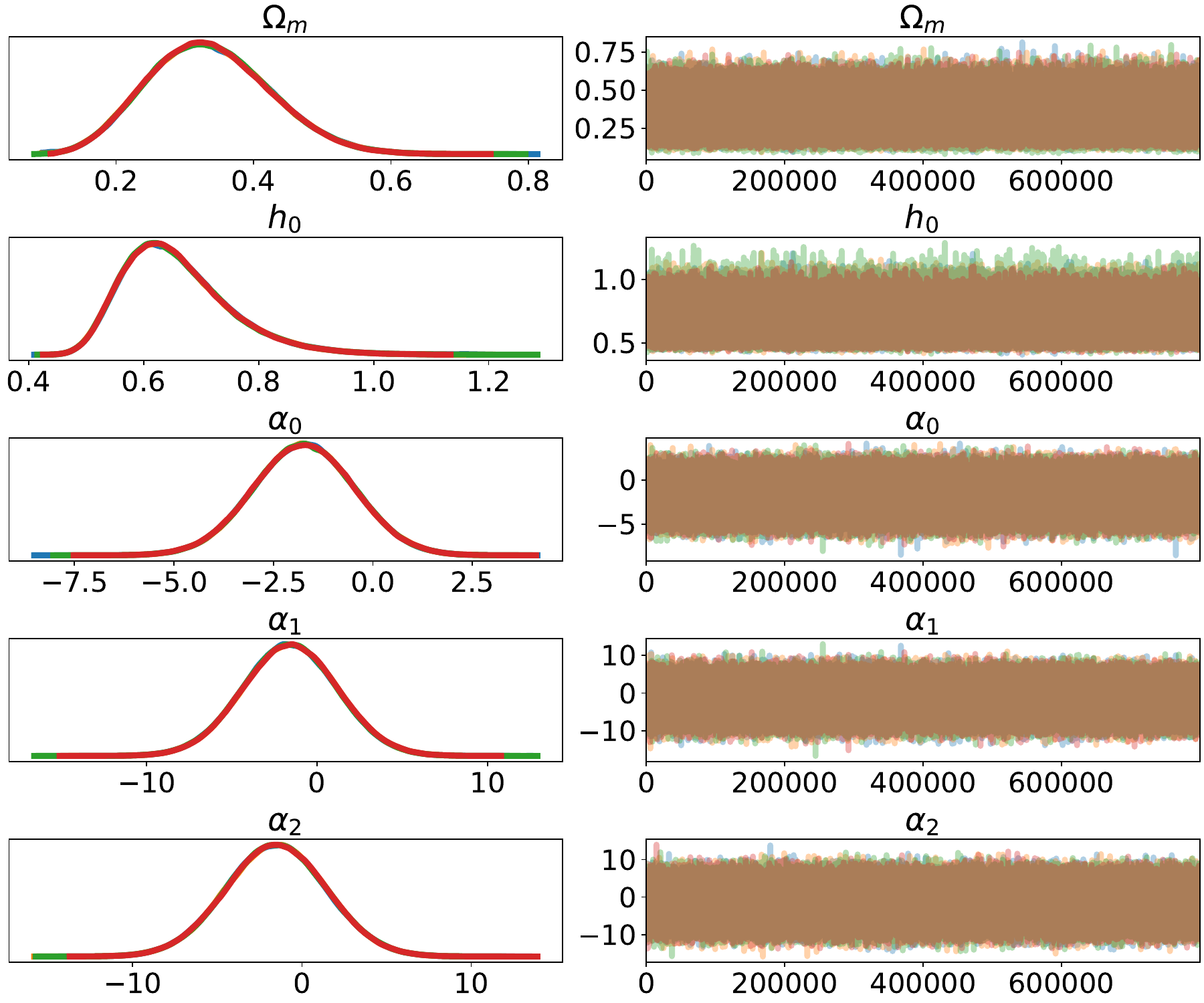}
	\caption{The column on the left consists of a smoothed histogram, 
	using kernel density estimation of the marginal posteriors of each stochastic random variable. 
	From top to bottom, the random variables are density parameters for matter $\Omega_m$, 
	reduced Hubble constant $h_0$, and the parameters of the equation of state of dark energy $\alpha_i$. 
	The right column contains the samples of the Markov chain plotted in sequential order.
	There are four MCMC chains that are plotted in different colors.
	These plots  are created with $(n_d, M_s) = (600, 800000)$ and all are for simulated Hubble parameter data-set. We use the same values of $n_d$ and $M_s$ in subsequent plots.} 
\label{fig:600_800000_simu_trace}
\end{figure*}

\begin{figure*}
	\includegraphics[scale=0.50]{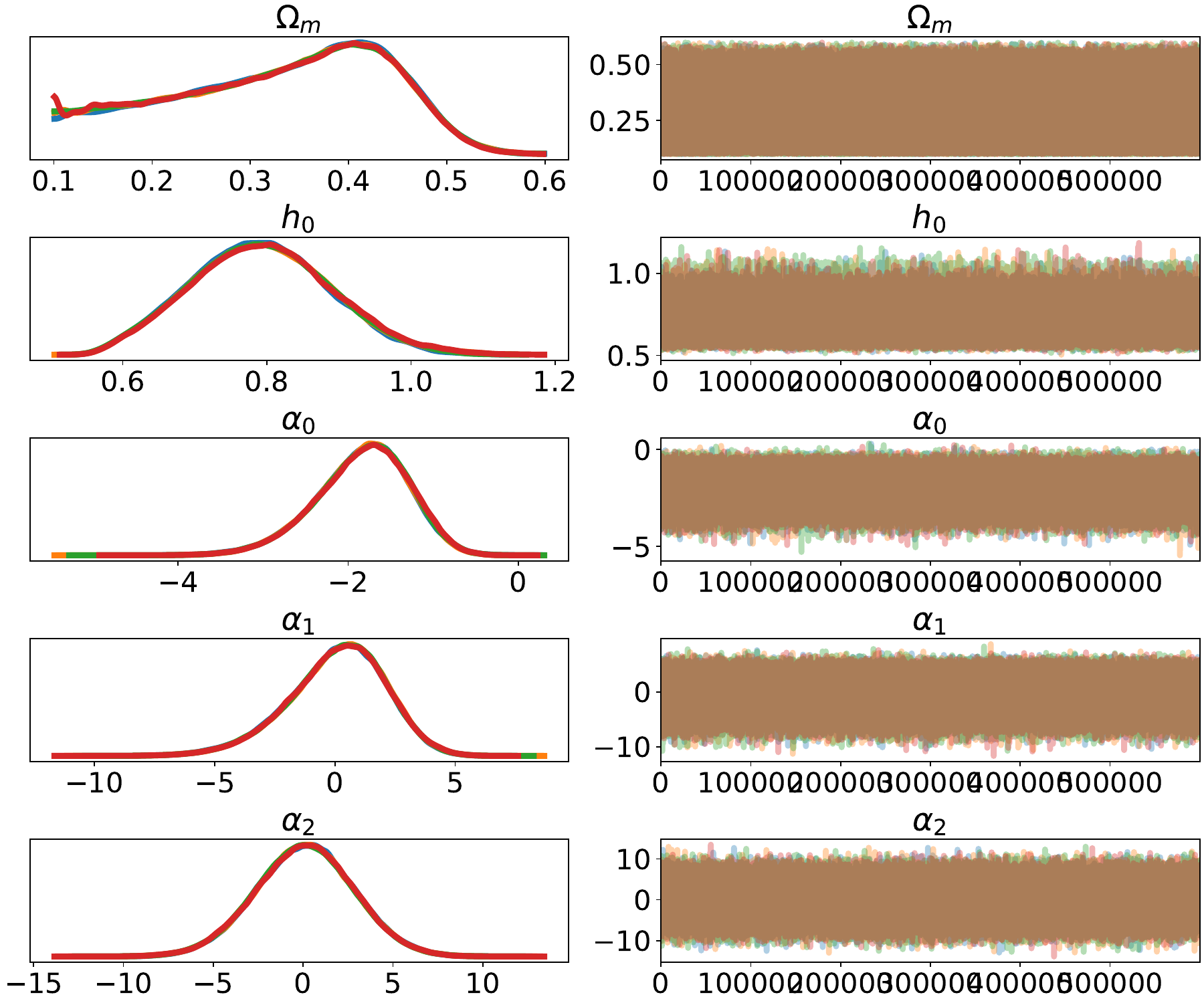}
	\caption{In this figure, the left column consists of a smoothed histogram, 
	using kernel density estimation of the marginal posteriors of each stochastic random variable. 
The parameters are same as those in fig(\ref{fig:600_800000_simu_trace}).
 The right column contains the samples of the Markov chain plotted in sequential order.
	These plots are also created with $(n_d, M_s) = (600, 800000)$ and all the plots are for observed Hubble parameter data-set.} 
\label{fig:600_800000_real_trace}
\end{figure*}

\begin{figure*}
	\includegraphics[scale=0.50]{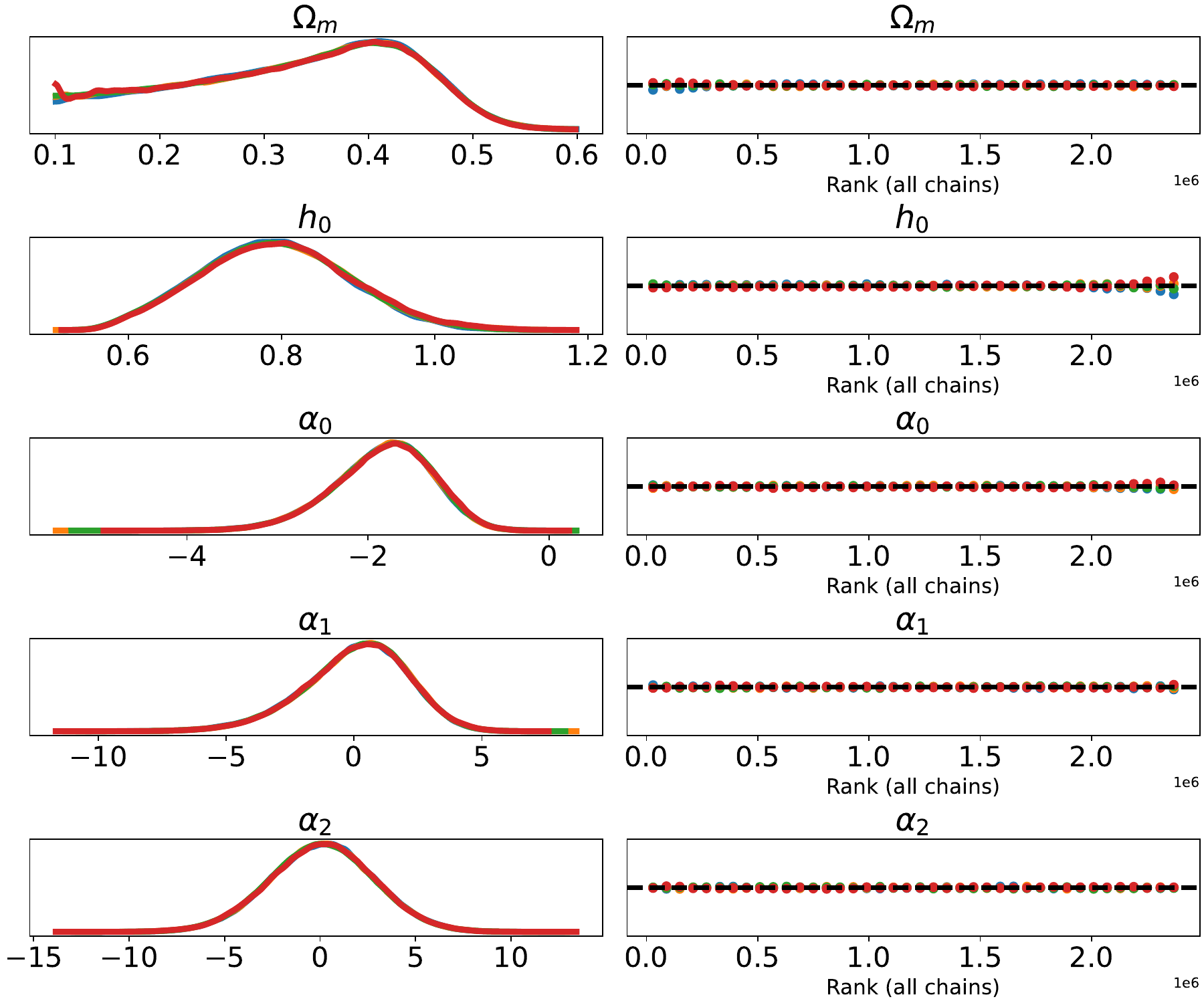}
	\caption{In this figure, the real Hubble parameter data-set is used to generate the plots. As in fig(\ref{fig:600_800000_simu_trace}) the left column consists of a smoothed histogram, 
	using kernel density estimation of the marginal posteriors of each stochastic random variable. 
	The right-hand column gives the auto-correlation for all the random variables for all the four independent chains, which are depicted with vertical lines.  
} 
\label{fig:600_800000_real_vlines}
\end{figure*}

Fig \ref{fig:600_800000_simu_trace}, \ref{fig:600_800000_simu_rank_bar}, and \ref{fig:600_800000_simu_vlines} shows the convergence of the MCMC chains for the simulated Hubble parameter data-set. 
These plots are created for $(n_d, M_s) = (600, 800000)$.
The convergence and the ability to draw successful samples from the parameter space is apparent for $(n_d, M_s) = (600, 800000)$ set in the case of Hubble parameter data-set.  
The same application for the real Hubble parameter data is shown in fig \ref{fig:600_800000_real_trace}, \ref{fig:600_800000_real_rank_bar}, and \ref{fig:600_800000_real_vlines}.
Fig(\ref{fig:600_800000_simu_vlines}, \ref{fig:600_800000_real_vlines}) gives the  auto-correlation 
of different parameters for different MCMC chains.  
We see from fig(\ref{fig:600_800000_simu_vlines}, \ref{fig:600_800000_real_vlines}) that auto-correlation of the parameters for different chains are zero.
For SNIa data-set also, the auto-correlation as well as the trace and rank-bar plots are similar,
and shows the convergence of the MCMC chain.


\section{Convergence plots for Metropolis-Hasting}
\label{MH_plots}
\begin{figure*}
        \includegraphics[scale=0.18]{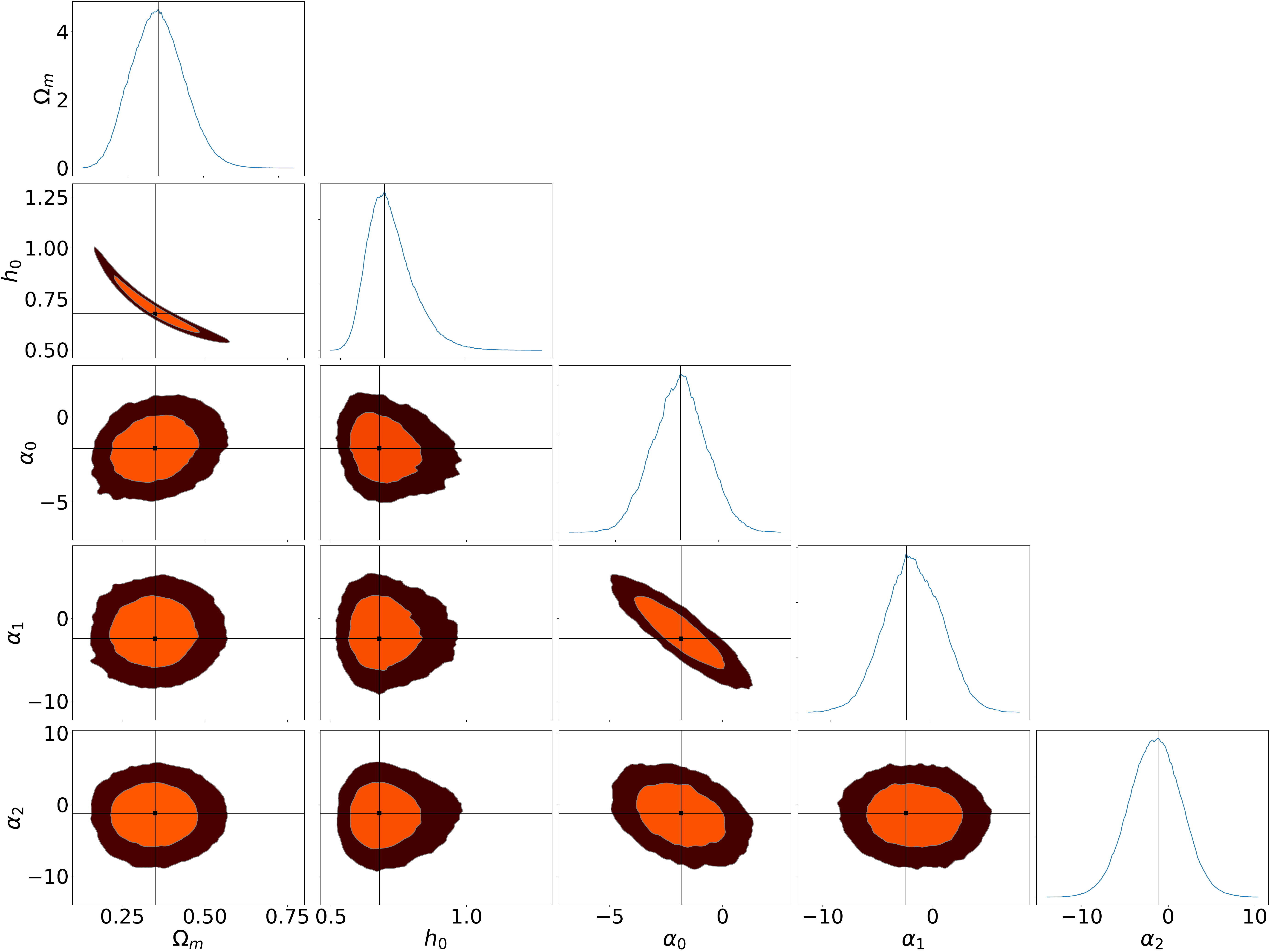}
    \caption{The plot shows the $1 \sigma$ and $2 \sigma$ contours for all the parameters along with their marginal probability density plots for the case of real data-set \citep{ohd1, ohd2, ohd3, ohd4, ohd5}.
    Constraints on the parameter is done using the Matropolis Hasting algorithm. 
The first plot of every column shows the marginal probability density plot, which give the maximum evidence of the real Hubble parameter data-set.}
\label{fig:MH_real_cont}
\end{figure*}

\begin{figure*}
	\includegraphics[scale=0.50]{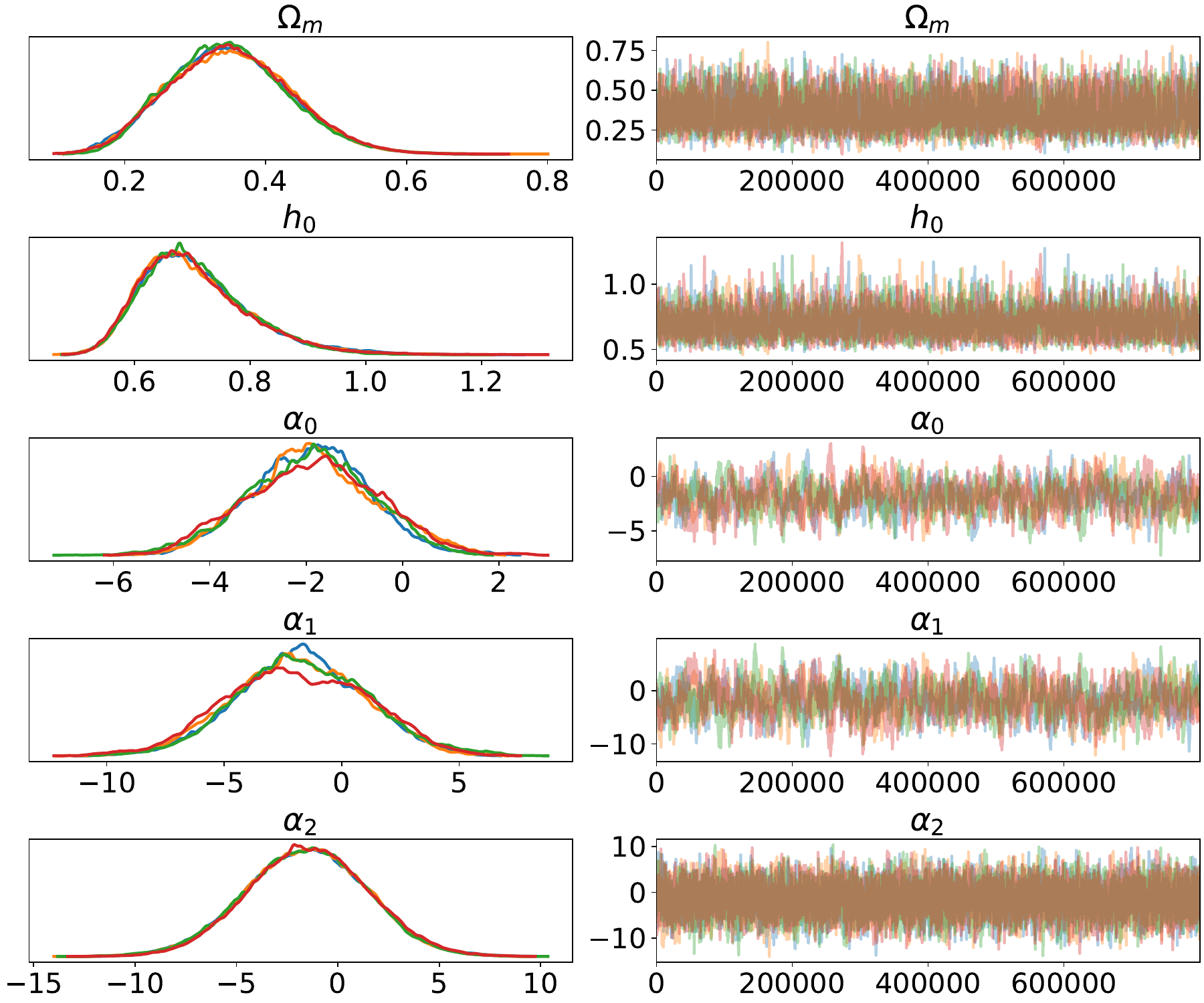}
	\caption{Here we use Matropolis-Hasting algorithm to explore the parameter space created for the model of eqn(\ref{eq:FRW}, \ref{eq:param_eos}).
	As in the fig(\ref{fig:600_800000_simu_trace}), the left consists of a smoothed histogram, using kernel density estimation of the marginal posteriors of each stochastic random variable. 
	These plots  are created with $(n_d, M_s) = (600, 800000)$ and all are for the real Hubble parameter data-set. 
	From top to bottom, the random variables are density parameters for matter $\Omega_m$, 
	reduced Hubble constant $h_0$, and the parameters of the equation of state of dark energy $\alpha_i$. 
	The right column contains the samples of the Markov chain plotted in sequential order.
	There are four MCMC chains that are plotted in different colors.} 
\label{fig:MH_600_800000_real_trace}
\end{figure*}

\begin{figure*}
	\includegraphics[scale=0.50]{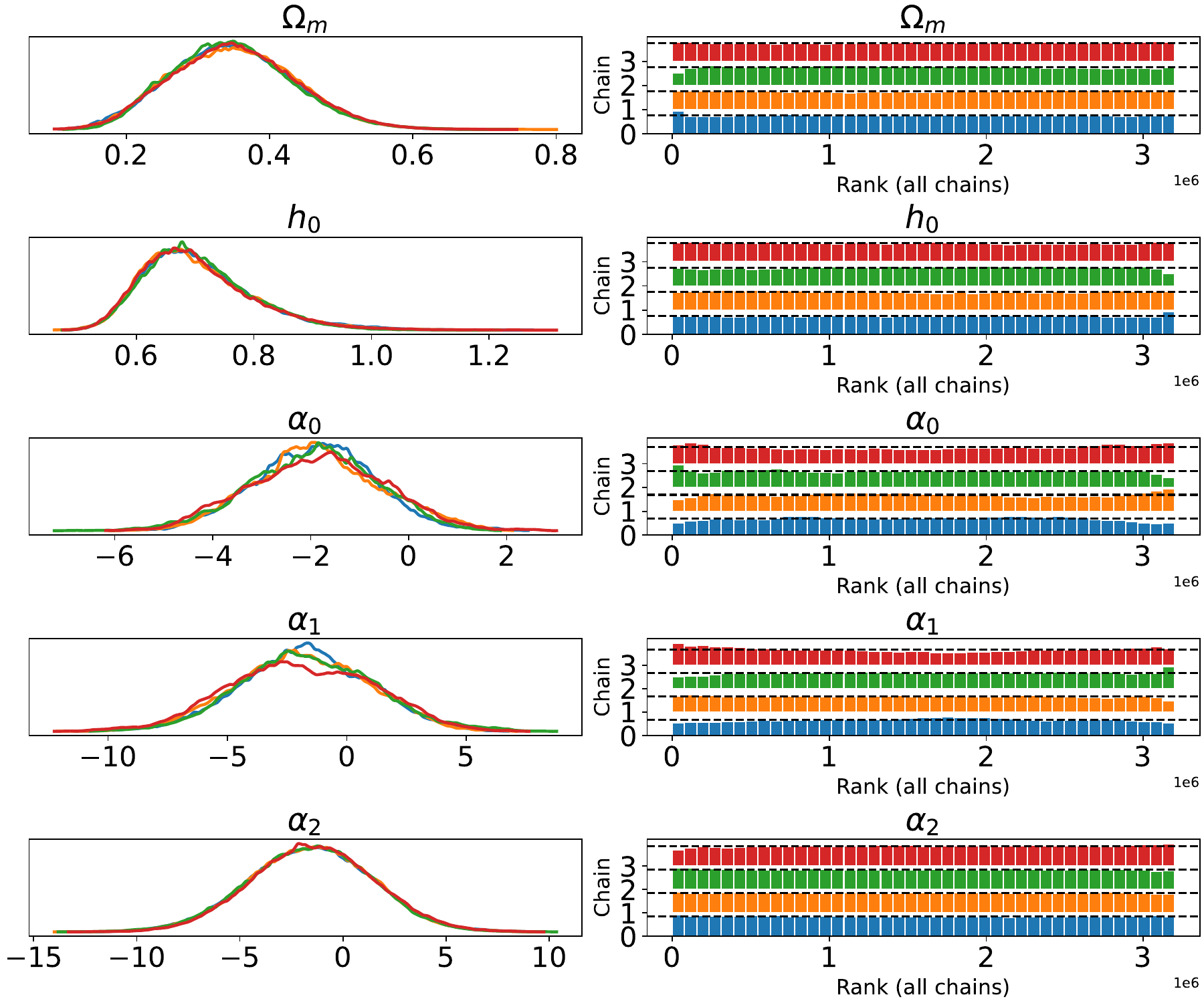}
	\caption{These plots are for the MH algorithm. 
	Smoothed histogram using kernel density estimation of the marginal posteriors of each stochastic random variable. 
Sequence of the parameters in the plot are same as in the fig(\ref{fig:MH_600_800000_real_trace}). 
	The four MCMC chains are plotted in different colors and all are independent of each other.
	} 
\label{fig:MH_600_800000_real_rank_bar}
\end{figure*}

\begin{figure*}
	\includegraphics[scale=0.50]{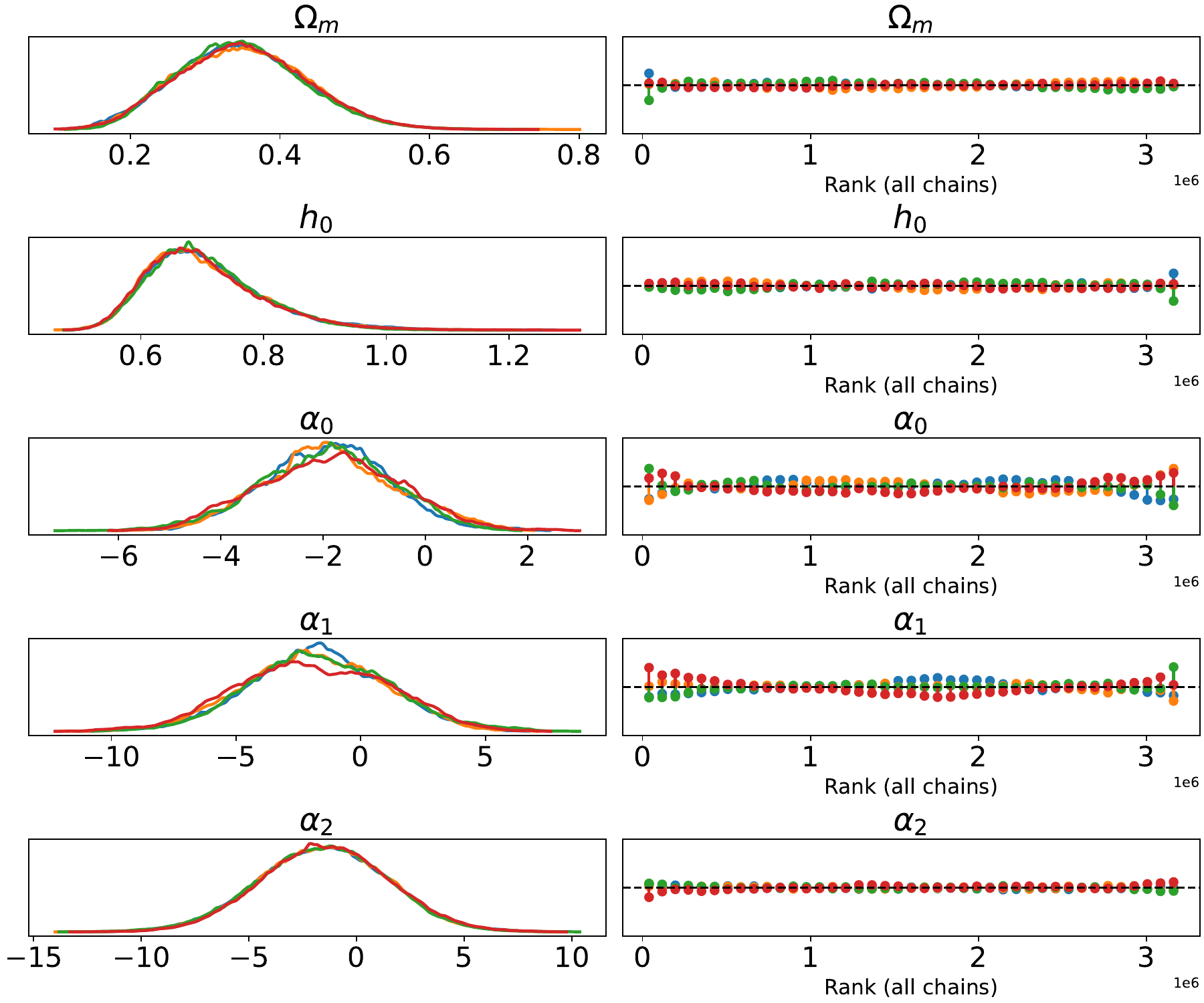}
	\caption{In this figure, the left column consists of a smoothed histogram, 
	using kernel density estimation of the marginal posteriors of each stochastic random variable. 
	The right-hand column gives the auto-correlation for all the random variables for all four independent chains, which are depicted with vertical lines.  
	As in the fig(\ref{fig:MH_600_800000_real_trace},\ref{fig:MH_600_800000_real_rank_bar}) these are also for the MH algorithm.} 
\label{fig:MH_600_800000_real_vlines}
\end{figure*}

For the comparison perpose we here give the trace-plot, rank-bar plot and vertical along with the contour plots for the real Hubble parameter data-set using Metropolis-Hasting algorithm (MH) fig(\ref{fig:MH_real_cont},\ref{fig:MH_600_800000_real_trace}, \ref{fig:MH_600_800000_real_rank_bar}, \ref{fig:MH_600_800000_real_vlines}). 
The number of data-points as well as the the sample points are same as in the NUTS algorithm. 
Markov chains are run for the simulated data-set. 
As we can see from the maximum evidence plots in the left of these figures, the convergence of the four chains along with their overlap is much less efficient than in the case of NUTS algorithm.
The the marginal posteriors of each stochastic random variables of the parameter space shows the same lake of convergence for the real data-set using MH. 
Convergence plot for the simulated data-set are also similar. 
For all the convergence plots the value of simulated data-set and the sample points for the MCMC chains are same as in the case of NUTS algorithm, $(n_d, M_s) = (600, 800000)$.
Figure \ref{fig:MH_real_cont} shows that for fix computational power and with the same values of $(n_d, M_s)$, NUTS constrains the parameter space in a better way than the MH. 

To compare the time taken by both NUTS and MH, we run the four chains of MCMC with model of eqn(\ref{eq:FRW},\ref{eq:param_eos}) and the real Hubble parameter data-set.
For different number of parameters we run the MCMC chains of MH and NUTS with a 64 bit x86\_ 64 architecture computer. 
The results are given in the table \ref{Table::Time_cpu}. 
We see that when the number of continuous parameter increases NUTS takes much smaller time then the traditional MH algorithm. 
We run the program with $(n_d, M_s) = (100, 100)$.  
These time periods are calculated as an average over the 20 identical runs.

When the number of continuous parameters increases NUTS works in a better way in comparison to MH. 
The fig(\ref{fig:MH_real_cont},\ref{fig:MH_600_800000_real_trace}, \ref{fig:MH_600_800000_real_rank_bar}, \ref{fig:MH_600_800000_real_vlines}) are clear indication to this. 
NUTS exploits the gradient information of the parameter space to achieve the must faster convergence in comparison to the MH. 
This can be clearly seen for the models with larger continuous parameter. 
\cite{pymc, gelman-rubin-1992, Hoffman_Gelman} provide a clear theoritical basis on which the effectiveness of the NUTS can be put above the MH. 

\begin{table}

\begin{tabular}{|l|*{2}{c|}}\hline
\backslashbox{param}{Algo}
&\makebox[3em]{MH}&\makebox[3em]{NUTS}\\\hline\hline
m = 3; $\Omega_m$, $h_0$ & 79.84 & 82.28\\\hline
m = 5; $\Omega_m$, $h_0$ & 207.20 & 85.561\\\hline
m = 10; $\Omega_m$, $h_0$ & 259.34 & 97.31\\\hline
\end{tabular}

\caption{Here, we give the total time taken by Matropolis-Hasting(MH) and NUTS algorithm. 
In the above table $m$ is the number of EoS parameter, eqn(\ref{eq:param_eos}).
Time taken is shown in the unit of seconds.
Each time is the average over 20 identical runs, with the 64 bit x86\_ 64 architecture computer.
$(n_d, M_s) = (100, 100)$ for all the runs.}
\label{Table::Time_cpu}
\end{table}


\pagebreak

\section*{Acknowledgements}
We would like to thank J S Bagla for useful suggestions. We would also like to thank PyMC3 team \citep{pymc, R:manual} for making their software open-source and user-friendly. 

\section*{Data Availability}
The observational data-set used in the analysis is publicly available and duly referred to in  the text \citep{ohd1,ohd2,ohd3,ohd4, ohd5, Scolnic:2021amr}.
The simulated data-set can be created by using the standard $\Lambda CDM$ model, using eqn(\ref{eq:FRW}-\ref{eq:mu}) of the text.
\vspace{-1em}

\pagebreak


\bibliography{references}




\end{document}